\documentclass[10pt]{article}
\usepackage{amssymb,color}
\usepackage{amsmath,graphicx,enumerate,textcomp}
\usepackage[abbrev]{amsrefs}
\usepackage[a4paper,top=1.in, bottom=1.5in, left=.8in, right=.8in, headsep=0.in, centering]{geometry}
\usepackage{amssymb,amsfonts,amsthm}
\usepackage{epsfig}
\usepackage{lipsum}
\usepackage{graphicx}
\usepackage{caption}
\numberwithin{equation}{section}

\newcommand{\eps}{\varepsilon}
\newcommand{\quotes}[1]{``#1''}

\title{An elementary model for an advancing autoignition front in laminar reactive co-flow jets injected into supercritical water. }

\author {Amanda Matson
\thanks{Department of Mathematical Sciences, 
Kent State University,
 Kent, OH 44242, USA. E-mail: {\tt amatson2@kent.edu  }}
 \and Michael C. Hicks
\thanks{ NASA Glenn Research Center, Cleveland, OH 44135, USA. E-mail: {\tt  michael.c.hicks@nasa.gov}}
\and  Uday G. Hegde
\thanks{
Department of Mechanical and Aerospace Engineering,
 Case Western Reserve University, Cleveland, OH 44106, USA.  E-mail: {\tt uday.g.hegde@nasa.gov}}
\and Peter  V. Gordon
\thanks{Department of Mathematical Sciences, 
Kent State University,
 Kent, OH 44242, USA. E-mail: {\tt gordon@math.kent.edu}}
}

\makeatother

\begin{document}

\maketitle
\begin{abstract} 
In this paper we formulate and analyze an elementary  model for the propagation of   advancing autoignition  fronts in reactive co-flow fuel/oxidizer  jets injected into an aqueous environment  at high pressure.
This work is motivated by the experimental studies of autoignition of hydrothermal flames performed at the high pressure laboratory of NASA Glenn Research Center.
Guided by  experimental observations,   we use several simplifying assumptions that allow the derivation of a  simple, still experimentally feasible, mathematical model for the propagation of advancing ignition fronts. The model consists of a  single diffusion-absorption-advection equation posed in an infinite cylindrical domain with a non-linear condition  on the boundary of the cylinder and  describes the temperature distribution within  the jet.
This model manifests an interplay of  thermal diffusion,  advection and volumetric  heat loss within a fuel jet which are balanced by the weak chemical reaction on the jet's boundary. 
We  analyze the model by means of asymptotic and numerical techniques and discuss feasible  regimes of propagation of  advancing ignition fronts.
In particular, we show that  in the most interesting  parametric regime
when the advancing ignition front is on the verge of extinction this model reduces to a one dimensional reaction-diffusion equation with bistable non-linearity. 
We hope that the present study will be helpful for the interpretation of existing experimental data and guiding of future experiments.
\bigskip
\begin{center}
{\bf Graphical Abstract}
\end{center}
\begin{figure}[h]
%\noindent \begin{raggedright}
\hspace*{-0.2cm}%
\begin{minipage}[c]{0.2\textwidth}%
\includegraphics[width=1.2in]{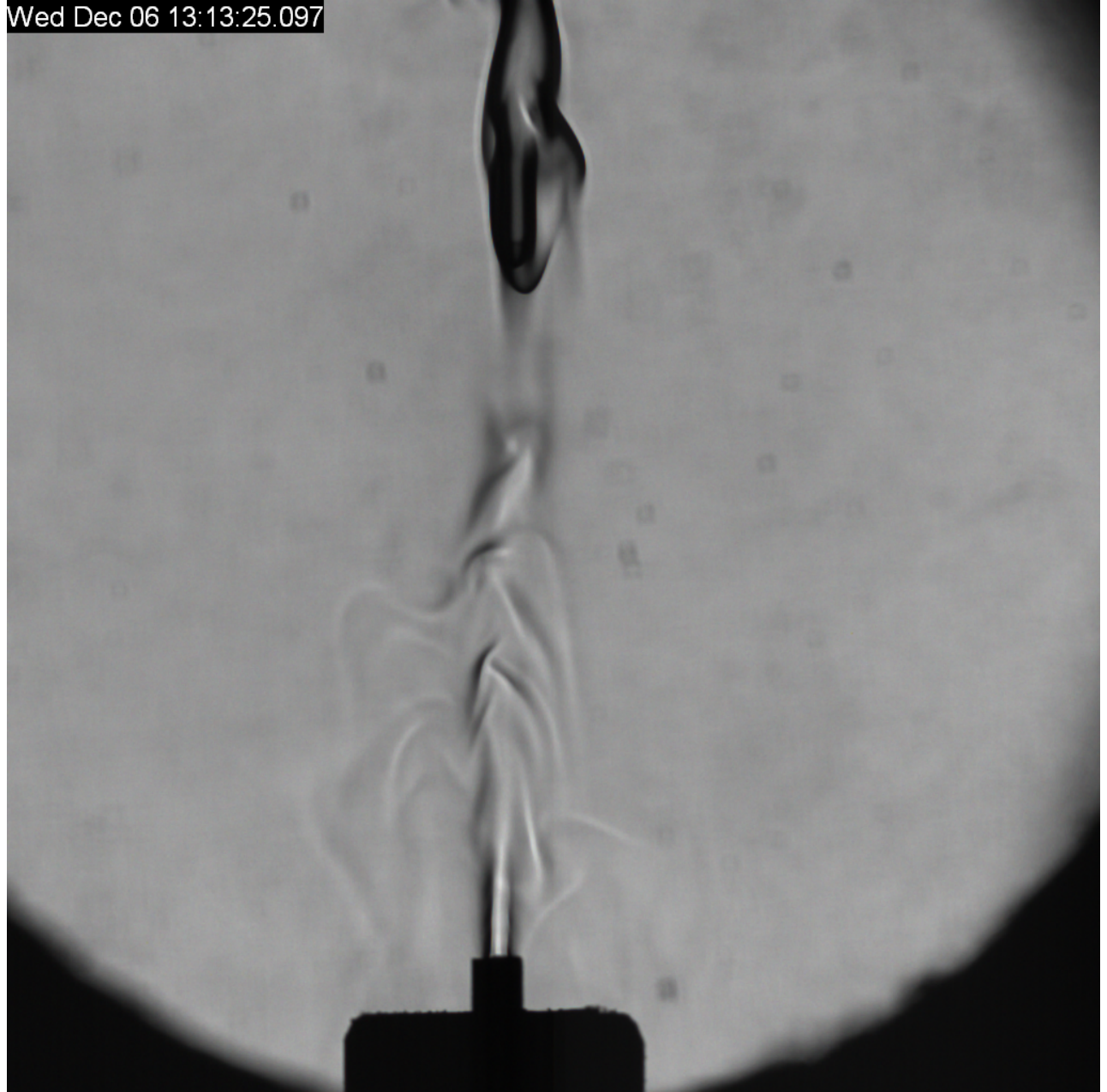} %
\end{minipage}\hspace*{-0.7cm} %
\begin{minipage}[c]{0.2\textwidth}%
\includegraphics[width=1.2in]{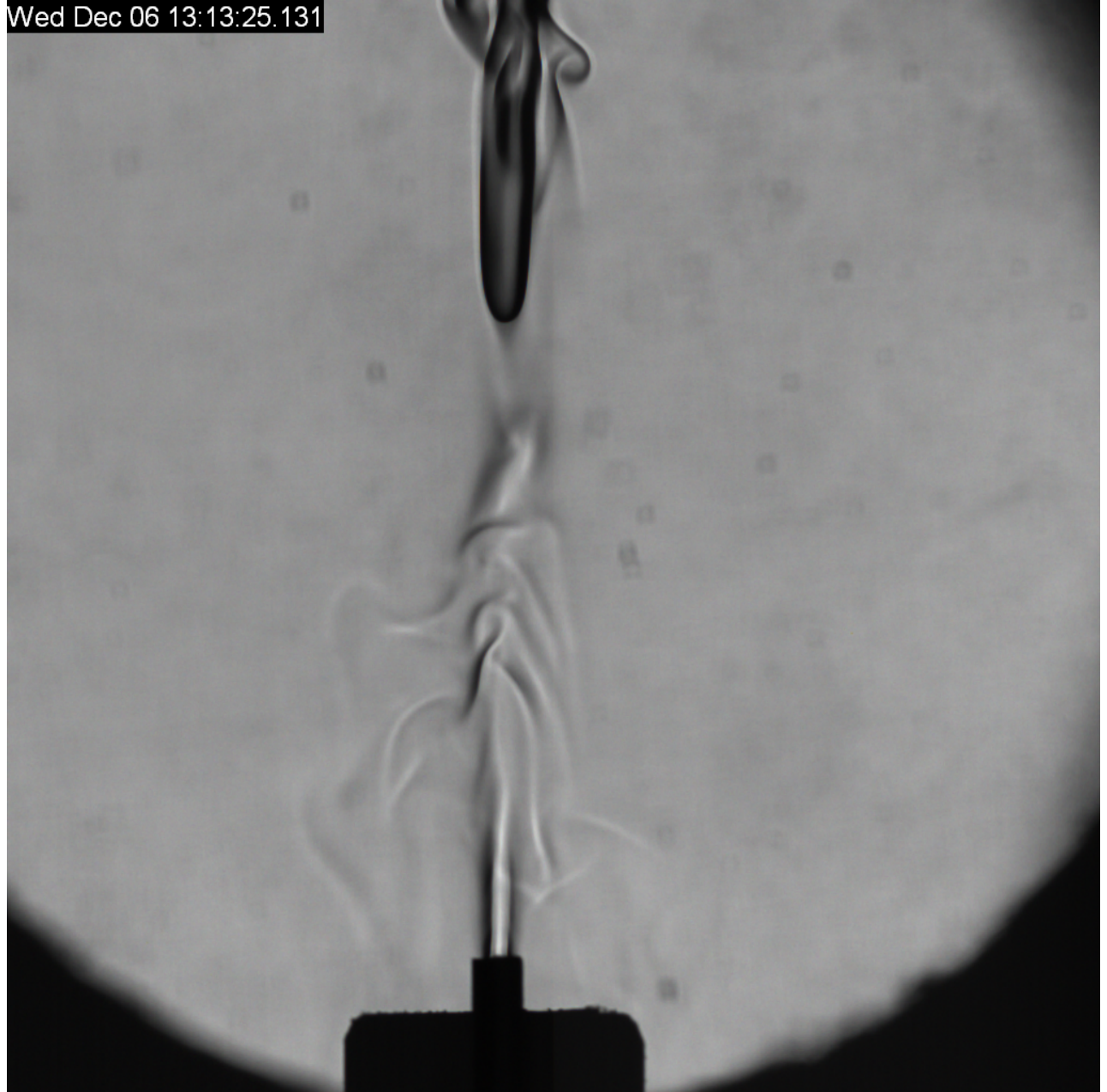} %
\end{minipage}\hspace*{-0.7cm} %
\begin{minipage}[c]{0.2\textwidth}%
\includegraphics[width=1.2in]{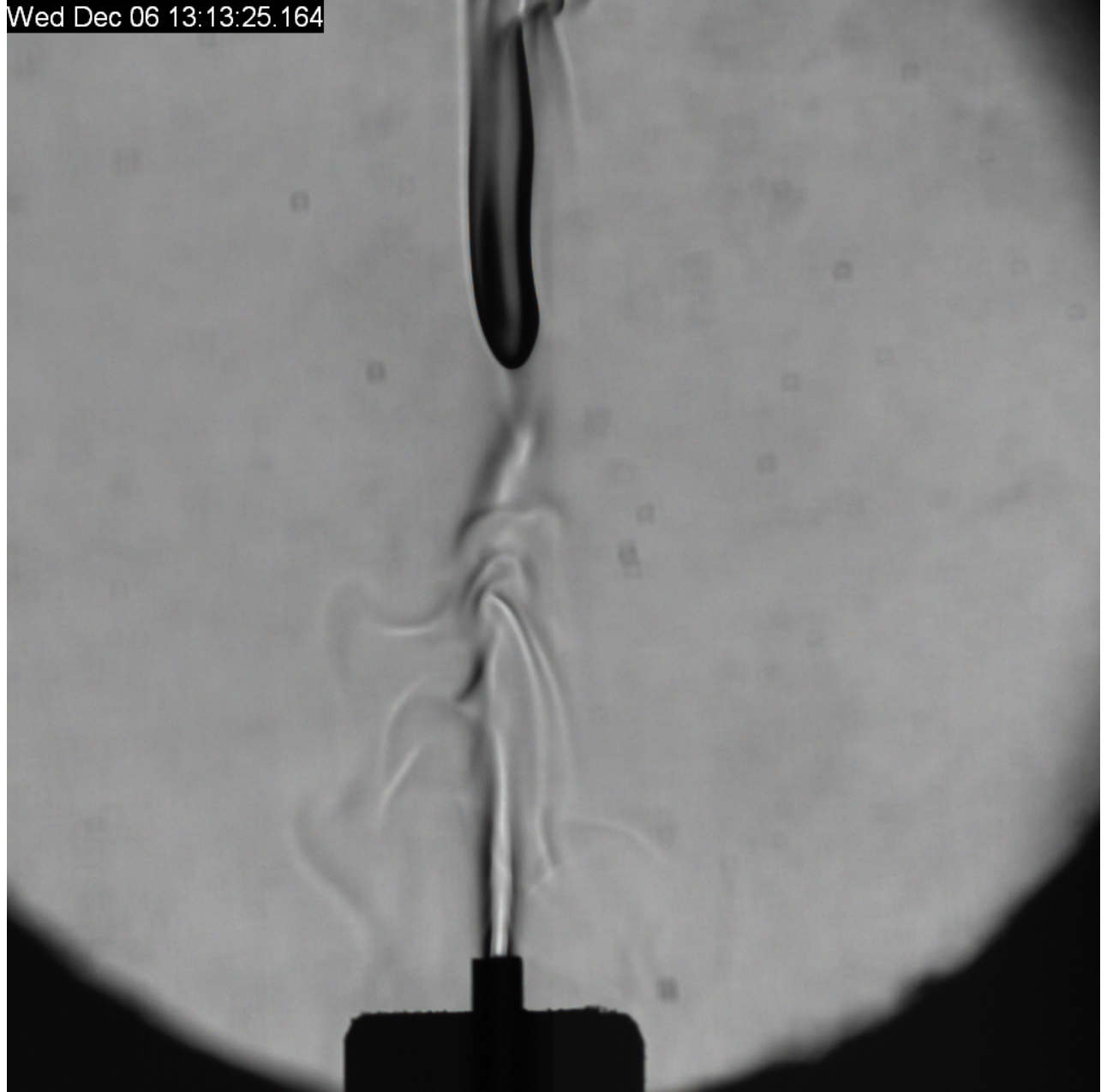} %
\end{minipage}\hspace*{-0.7cm} %
%\vspace*{\bigskipamount}
%\par\end{raggedright}
%\noindent \begin{raggedright}
%\hspace*{-0.5cm}%
\begin{minipage}[c]{0.2\textwidth}%
\includegraphics[width=1.2in]{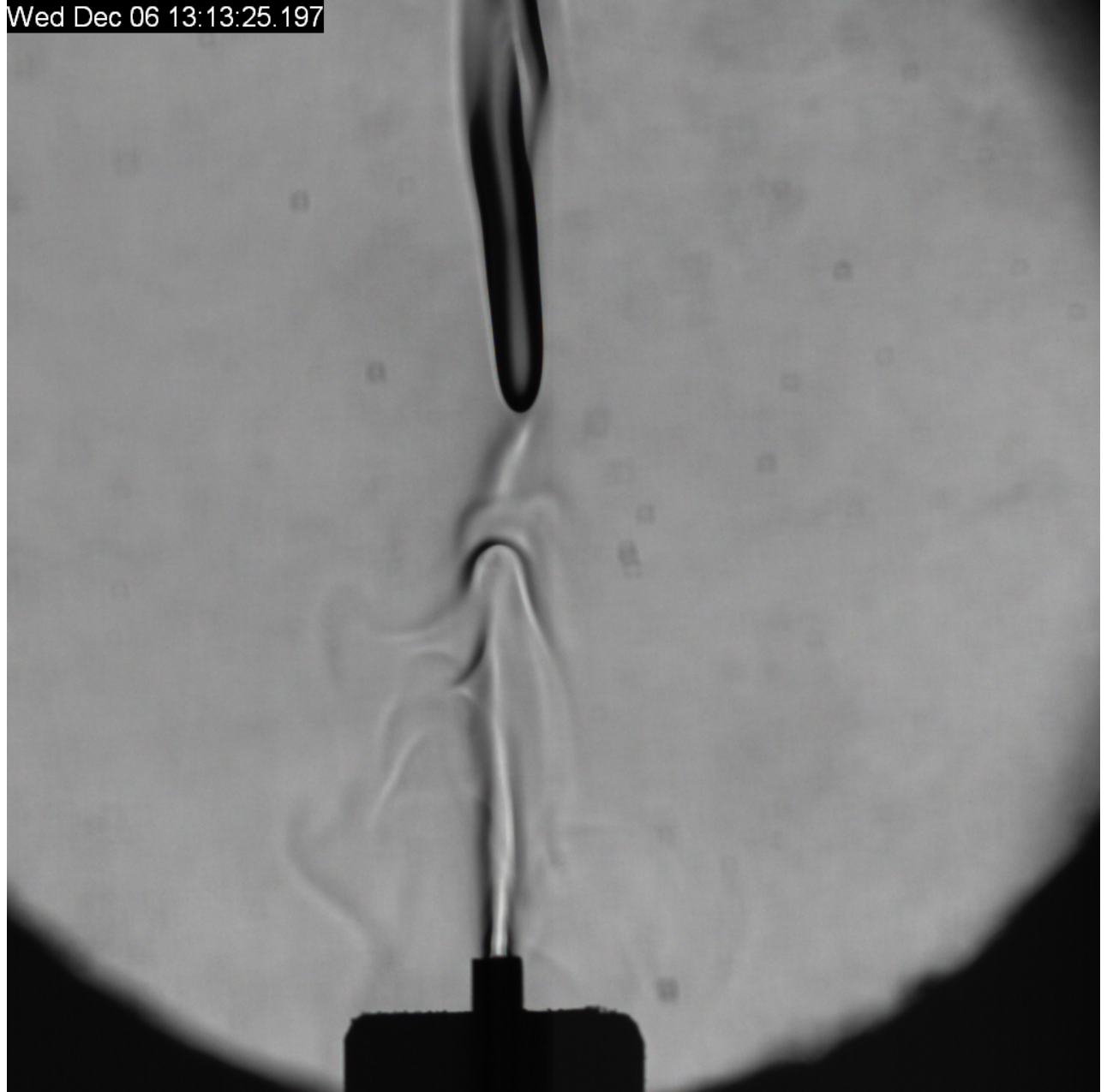} %
\end{minipage}\hspace*{-0.7cm} %
\begin{minipage}[c]{0.2\textwidth}%
\includegraphics[width=1.2in]{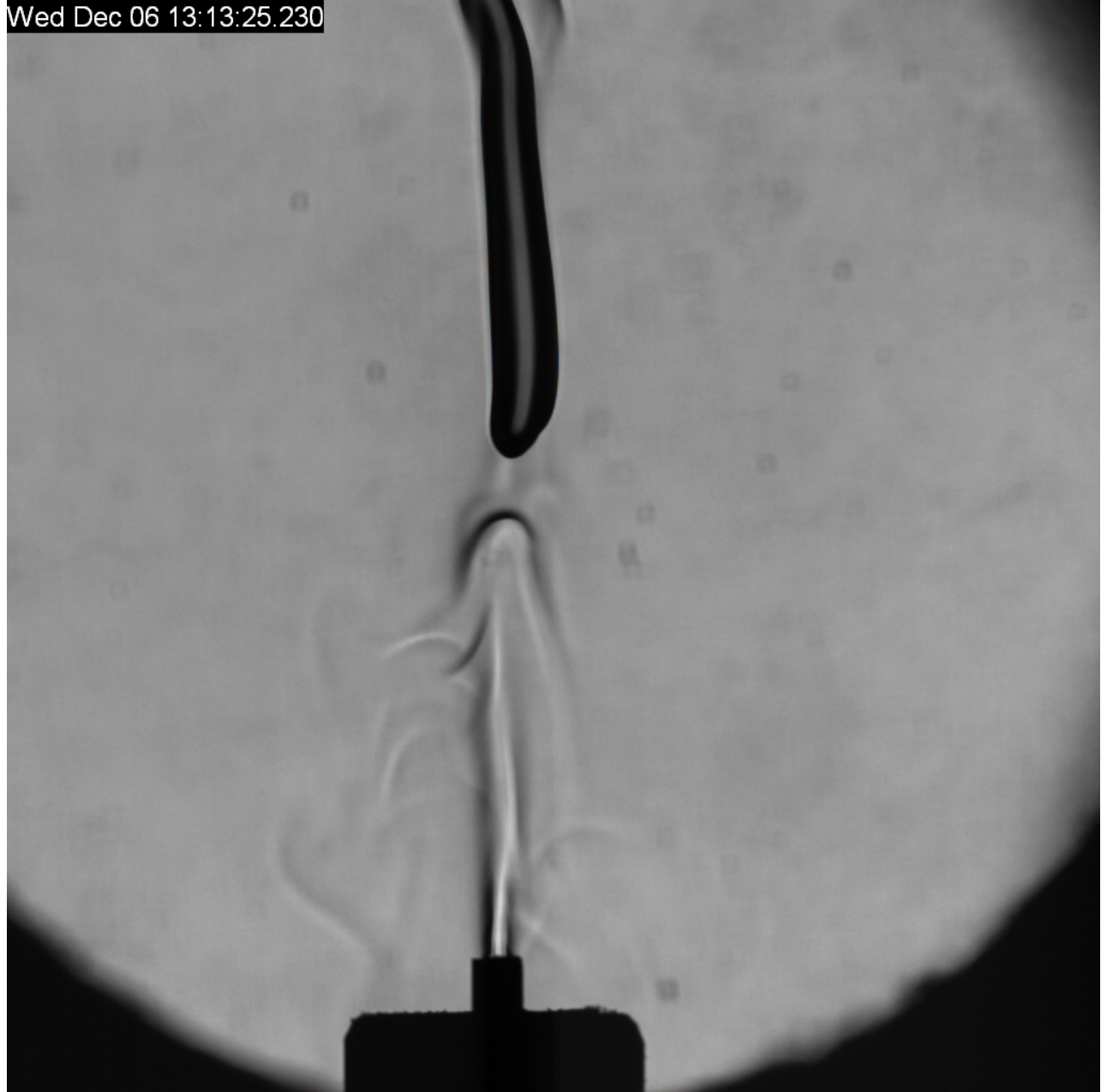} %
\end{minipage}\hspace*{-0.7cm} %
\begin{minipage}[c]{0.2\textwidth}%
\includegraphics[width=1.2in]{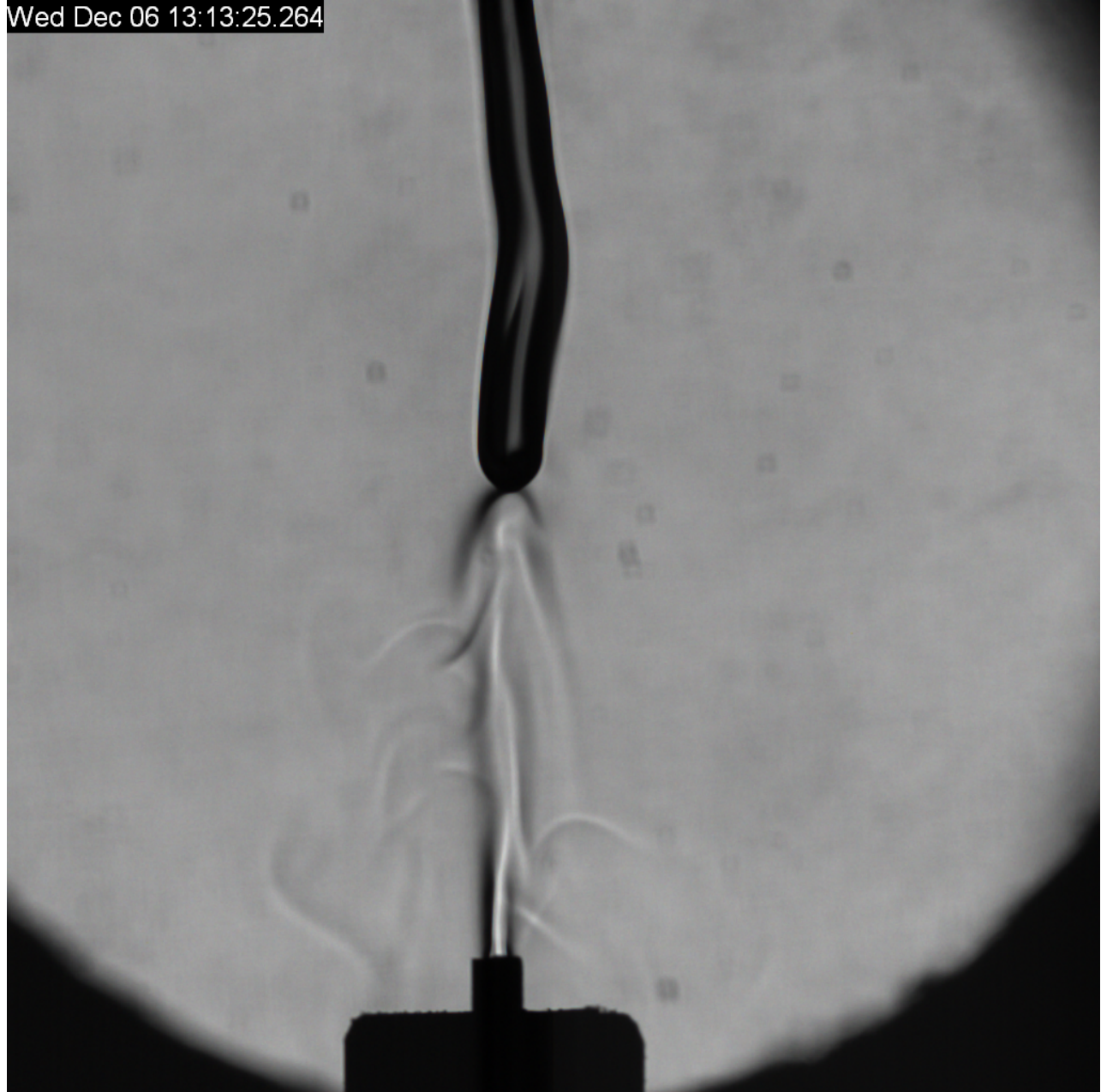} %
\end{minipage}
%\vspace*{\smallskipamount}
%\par\end{raggedright}
%\noindent \raggedright{}
\caption*{ Advancing autoignition front in a co-flow reactive laminar jet injected into supercritical water.}
%\label{fig:2} 
\end{figure}

\end{abstract}

\bigskip


\medskip

\section{Introduction}

Supercritical water oxidation (SCWO) is an advanced technology  for the purification of organically contaminated aqueous waste streams. 
This technology is based on the fact that organic waste can be efficiently oxidized in water above its critical point ($374 {\rm C^o}, 22 {\rm MPa}$).
Indeed, at supercritical conditions organic compounds and gases become fully soluble in water which results in high reaction rates between dissolved 
oxygen and organic materials and leads to almost prefect conversion \cite{intro1,intro2}. SCWO is often quoted as  "green"  technology as no
hazardous pollutants such as $NO_x$ and $SO_x$ are produced in the process. These advantages makes SCWO very promising 
in application to waste management both on earth and in space. 
In early designs of SCWO reactors, the oxidation process was taking place at relatively low temperatures. More recently, however, a number of SCWO
reactors were designed such that  the oxidation results in ignition and formation of controlled {\it hydrothermal flames}  which increase conversion
efficiency and reduce required residence time and hence  viewed as beneficent (see e.g  \cite{intro3,intro4,intro5,intro6}).

Hydrothermal flames were first experimentally observed
by W. Schilling and E.U. Franck \cite{SF}  in the late 1980's  and  since that time were extensively studied,
see reviews \cite{MikeRev,Rev1} for more details. We also note that beside its clear  technological relevance hydrothermal flames serve as a canonical system for studying
combustion at highly elevated pressures.

Experimental studies of hydrothermal flames are time consuming and rather expensive.
Therefore,
the theoretical  understanding of principal mechanisms and parameters  controlling  ignition and formation of hydrothermal flames is of importance for
guiding experiments and interpreting experimental observations. Modeling of hydrothermal flames in full generality is a challenging problem that requires solving an extremely complicated system of equations that describe multiple 
chemical reactions taking place in hydrothermal flames coupled with highly nontrivial hydrodynamic equations and equations of state.  Such models involve a large number
of various physical, chemical and geometric parameters some of which are difficult to estimate which presents an additional challenge in modeling.
As noted in a plenary lecture of G.I.  Sivashinsky
(one of the top experts in the modern combustion theory) at the  29th  international symposium on combustion in 2002 \cite{Grisha_pl}, there are basically two approaches for modeling combustion systems:
\quotes{ {\it When modeling a combustion system, one may try
to include everything that is likely to be of quantitative importance, or one may intentionally ignore
certain aspects to elucidate the impact of those that
are retained. The first approach is necessary if the
goal is to obtain numbers for comparison with experimental measurements or for the design of practical devices. The second approach is of great value
when the goal is to gain physical insight by making
the problem tractable.}}
In this paper we adopt the second approach and derive a minimal, mathematically tractable, still experimentally feasible  model
 for the initial stages of the reactive process in laminar reactive jets injected into supercritical water. Despite its limitations, the model  allows us to understand the  role of 
several principal geometric and physicochemical parameters that determines the reactive process at these stages.

The present paper deals with modeling   {\it autoignition}, a rapid spontaneous  oxidation of reactive matter being initially in a non-reactive state.
  Autoignition  is one of the most basic phenomena studied in combustion theory. The experimental and theoretical explorations of autoignition have  a long and rich history.  Mathematical modeling of autoignition processes  traces back 
 to classical works of N.N. Semenov, D.A.  Frank-Kamnetskii and Ya. B. Zeldovich
in the 1920s and 1930s \cite{Sem,FK,ZBLM}.  The Semenov-Frank-Kamenetskii approach to autoignition, among other factors,  utilizes the fact that  the  chemical reaction in reactive media 
prior to autoignition is relatively weak.  Therefore, the  fuel
and oxidizer concentration fields as well as the hydrodynamics of the reactive media can be viewed as independent of the chemical reaction and hence regarded as given.
 This  leads to dramatic simplifications of governing equations  and allows  the  derivation of   mathematically tractable  models that can be analyzed in full detail.
 This, in turn, allows us to obtain a detailed picture of how autoignition scenarios depend on principal  physicochemical and geometric parameters of the specific problem.
 The general Semenov-Frank-Kamenetskii approach was adapted for studying autoignition in various reactive systems on different levels of complexity and proved to be a very effective 
 tool \cite{Law}.

The homogeneous autoignition described by the classical  Semenov's  theory may take place in some uniformly mixed  system well isolated from the surroundings, but this situation is  rather untypical.
It is much more common when autoignition is initiated in a strongly localized region called the {\it ignition kernel}. The formation of the localized  autoignition kernel, in the classical
setting,  is commonly modeled using the non-steady Frank-Kamenetski theory of thermal explosion. In this case, the formation of the ignition kernel is associated with a 
formation of a point singularity in the underlying system of governing equations.  It is important to note that direct application of the Semenov-Frank-Kamenetskii approach is
limited by its assumptions and generally provides an adequate description of the oxidation process up to the point when the ignition kernel is formed. Indeed,  in many combustion systems the formation of
 the ignition kernel immediately triggers strong chemical reaction and results in  instantaneous flame propagation. In such systems, the post ignition behavior
 cannot be possibly described in a framework of the ignition theory.
 However,  in certain reactive systems, 
 the ignition kernel may expand through the reactive media forming an {\it advancing ignition front,} and the visible flame appears only after some induction time.
  Let us note that the chemical reactions in the ignition fronts, in contrast to flame fronts,  is relatively weak
  and results in relatively small elevations of temperature and negligible consumption of the reactants. In addition, ignition fronts do not significantly disturb the underlying
 hydrodynamics picture. Hence, principle assumptions of the Semenov-Frank-Kamenetskii theory remain valid and can be utilized for modeling ignition fronts.

\begin{figure}[h]
\noindent \begin{raggedright}
\hspace*{-0.1cm}%
\begin{minipage}[c]{0.2\textwidth}%
\includegraphics[width=1.25in]{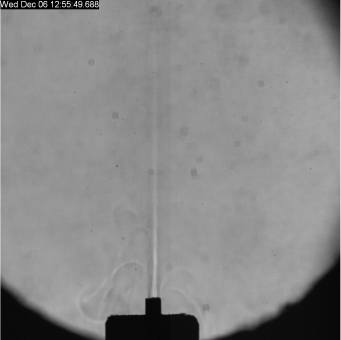} %
\end{minipage}\hspace*{-0.1cm} %
\begin{minipage}[c]{0.2\textwidth}%
\includegraphics[width=1.25in]{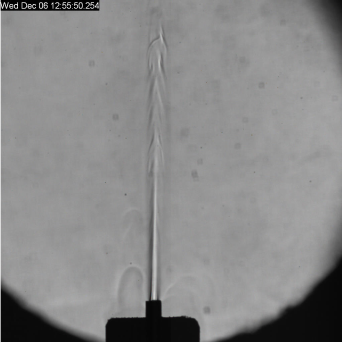} %
\end{minipage}\hspace*{-0.1cm} %
\begin{minipage}[c]{0.2\textwidth}%
\includegraphics[width=1.25in]{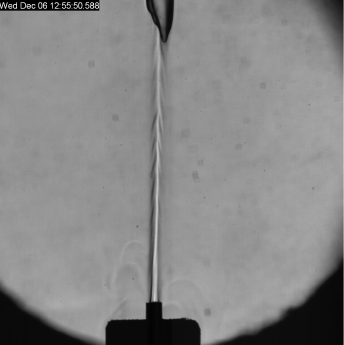} %
\end{minipage}\hspace*{-0.1cm} 
%\vspace*{\bigskipamount}
%\par\end{raggedright}
%\noindent \begin{raggedright}
%\hspace*{0.5cm}%
\begin{minipage}[c]{0.2\textwidth}%
\includegraphics[width=1.25in]{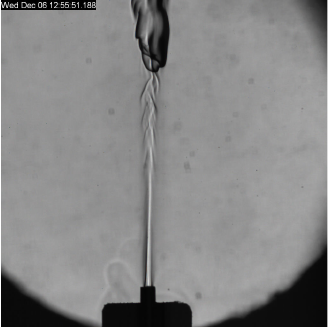} %
\end{minipage}\hspace*{-0.1cm} %
\begin{minipage}[c]{0.2\textwidth}%
\includegraphics[width=1.25in]{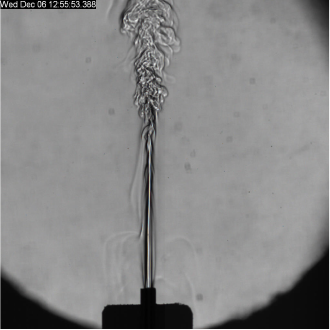} %
\end{minipage}\hspace*{-0.1cm} %
\par\end{raggedright}
\noindent \raggedright{}\caption{{Series of  shadowgraph   images of the combustion vessel
showing:  established laminar cylindrical co-flow  jet (first image), co-flow jet right prior to formation of the ignition kernel (second image),
 spontaneous formation and development of an ignition kernel (third and fourth image) and slightly turbulent  jet after the extinction (fifth image).
}}
\label{fig:1} 
\end{figure}

\begin{figure}[h]
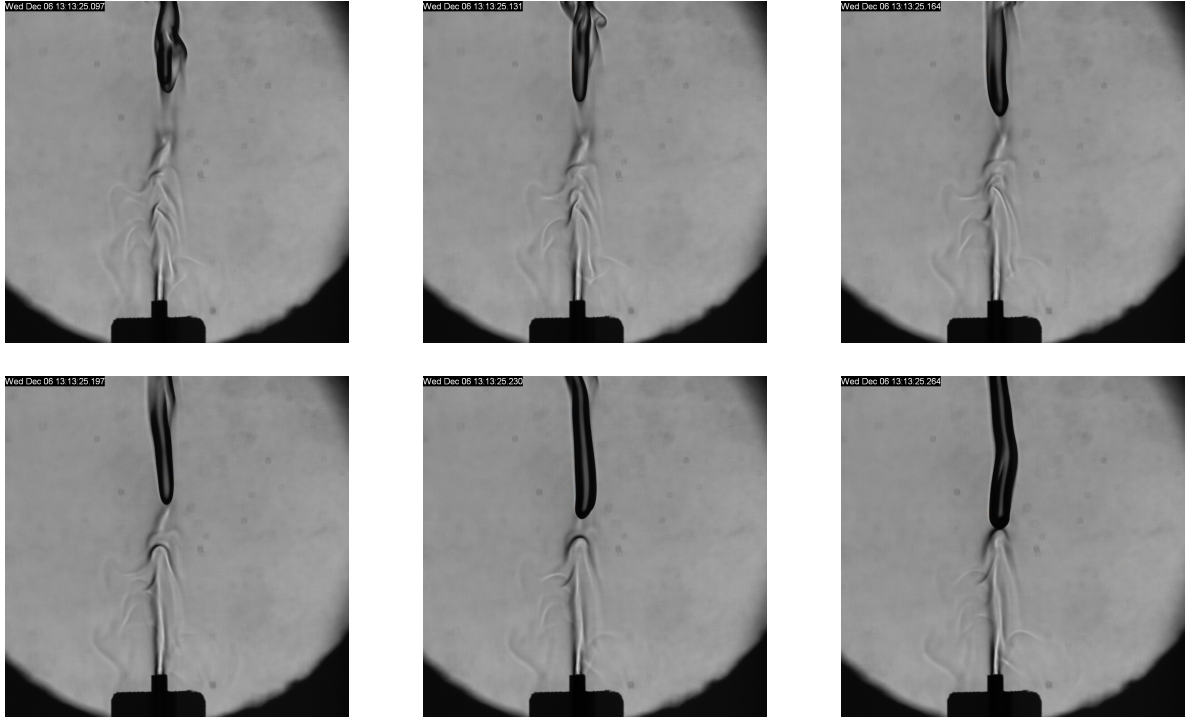

\noindent \begin{raggedright}
\hspace*{-0.5cm}%
\begin{minipage}[c]{0.3\textwidth}%
\includegraphics[width=2.3in]{fig2a} %
\end{minipage}\hspace*{0.3cm} %
\begin{minipage}[c]{0.3\textwidth}%
\includegraphics[width=2.3in]{fig2b} %
\end{minipage}\hspace*{0.3cm} %
\begin{minipage}[c]{0.3\textwidth}%
\includegraphics[width=2.3in]{fig2c} %
\end{minipage}\vspace*{\bigskipamount}
\par\end{raggedright}
\noindent \begin{raggedright}
\hspace*{-0.5cm}%
\begin{minipage}[c]{0.3\textwidth}%
\includegraphics[width=2.3in]{fig2d} %
\end{minipage}\hspace*{0.3cm} %
\begin{minipage}[c]{0.3\textwidth}%
\includegraphics[width=2.3in]{fig2e} %
\end{minipage}\hspace*{0.3cm} %
\begin{minipage}[c]{0.3\textwidth}%
\includegraphics[width=2.3in]{fig2f} %
\end{minipage}\vspace*{\smallskipamount}
\par\end{raggedright}
\noindent \raggedright{}\caption{{Series of consecutive shadowgraph images of the combustion vessel
showing a propagating traveling front of ignition. The snapshots were
taken at time instances about $35~ ms$ apart. The images were taken
in the high pressure combustion laboratory at the NASA Glenn Research
Center.}}
\label{fig:2} 
\end{figure}

 The formation and propagation of the advancing  ignition fronts were  observed in experimental studies of hydrothermal flames at the high pressure laboratory at
 NASA Glenn Research Center.  In these studies,  the co-flow jets consisting of  an inner jet of fuel (ethanol heavily diluted with  water and ethanol volume fraction as low as $20\%$) and outer  jet of an oxidizer (air)  were injected into a highly pressurized combustion vessel filled with water at supercritical 
 state. The experimental apparatus and  detailed description of the experimental procedure can be found in \cite{exp}. It was reported in \cite{exp} that at relatively slow injection rates  ($\sim 1-2 ~{\rm ml/min}$)  the injected substances formed  predominantly laminar jets of cylindrical shape with a sharp boundary between the  inner and outer parts.
 It was observed that, under  appropriate conditions, an ignition kernel is spontaneously formed in such jets  at  a certain  position that is  well elevated from the injection point. 
 Our earlier theoretical studies  of autoignition of laminar reactive jets \cite{Jet1,Jet2} were targeted to identifying principal physicochemical and geometric parameters
 that determine whether the formation of the ignition kernel  takes place. The results of the analysis of these models are in  excellent  quantitative  agreement with experimental observations.
 Moreover, as was shown in \cite{kita}  the predictions of a model discussed in \cite{Jet1} with certain adjustments of parameters are quite in line with the results of numerical simulations of a detailed  model for hydrothermal flames.
  Let us note that the formation of an ignition kernel  by no means guarantees the successful autoignition of the jet. Indeed, the  ignition kernel may be blown off
 by the advection and be eventually extinct  as shown in Figure \ref{fig:1}. 
 % %%%%%% First experimental figure
Hence, formation of the ignition kernel is a necessary, but not  sufficient condition for the successful ignition
 of the hydrothermal flame. As evident from studies presented in \cite{exp} in the case of successful ignition, the formation of the ignition kernel is followed by  the development  of a sausage-like 
 structure of weakly reactive substance at slightly elevated temperature which propagates  toward the injection inlet and eventually ignites the jet (see Figure \ref{fig:2}).
 We refer to this structure as an advancing ignition front.
In \cite{Jet4} we proposed a model for propagation of the ignition fronts. In that model, we assumed that the physicochemical characteristics of both the inner jet of fuel
 and outer jet of oxidizer are identical.  We also assumed that the volumetric  heat loss within both parts of the co-flow jet  are  the same and depends linearly on temperature.
 Finally, the  stepwise temperature kinetics  for the chemical reaction taking place on the boundary between the  inner and outer parts of the jet was adapted. The model presented in \cite{Jet4} reproduced certain experimentally observed
 effects such as propagation and blow off of the ignition front. Moreover, the closed form expression of the velocity  of the ignition front obtained in \cite{Jet4}
 allowed the formulation of the critical condition that ensures downstream propagation of an ignition front. The analysis of this critical condition showed how parameters of the problem
(ignition temperature, thermal conductivity, injection velocity, heat loss parameter, reaction intensity and radius of the jet) influence propagation regimes.
The results of this analysis were  in line with experimental observations. While the model presented in \cite{Jet4} provided some insight on how and when the ignition
front propagates,  some modifications of that model  are desirable to gain further understanding of the phenomena. Specifically, it is of interest to consider
a situation in which the  physicochemical characteristics of the inner and outer jets
  differ and the reaction and heat loss mechanisms are of a more general form. 
  Indeed, in  operating regimes of interest, the  thermal diffusivity of the fuel/water inner jet  is about twice as large as the oxidizing outer jet \cite{nist}.
 Moreover, the volumetric heat loss, mostly attributed to water \cite{hl} in the case of weak reactions, is substantially larger in the fuel/water part of the jet in comparison to  the one
 in the oxidizer. In this paper we propose  a model for the propagation of an advancing ignition front 
  that takes into account the factors discussed above. Surprisingly enough, the model derived 
 in this paper is, in many ways, simpler than the one proposed in \cite{Jet4}. 
 The current study is  a continuation of  our theoretical and experimental works \cite{Jet1,Jet2,Jet3,Jet4,GMN2020,exp,exp1,exp2}  targeted to understand   principal physical mechanisms controlling  oxidation and autoignition  in reactive jets.

The paper is organized as follows. In section \ref{s:model}, we state main assumptions of our theory and derive a mathematical model for studying advancing fronts of autoignition.
This model constitutes a single diffusion-absorption-advection equation posed in a cylindrical domain with non-linear boundary conditions. This problem describes an interplay of advection,
diffusion and heat loss within the fuel part of the co-flow jet that balance heat production of its boundary and connect reactive and non-reactive states far behind and far ahead of the ignition front.
We then state the critical condition that guarantees successful autoignition of the reactive jet. In section \ref{s:analysis}, we present the analysis of the derived model
and discuss various regimes of propagation. In the last section, we summarize the results obtained in this paper.

\section{Derivation of the model}\label{s:model}
\begin{figure}[h]
\centering \includegraphics[width=3in]{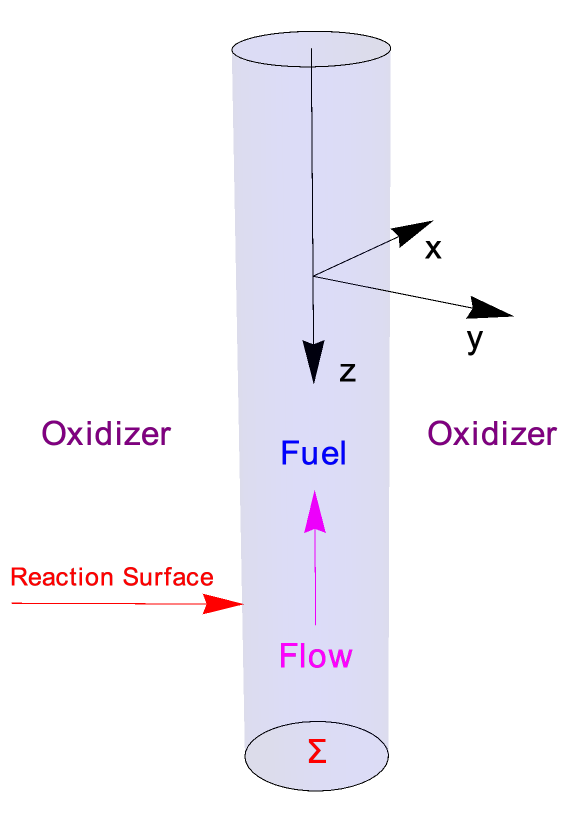} 
\caption{Sketch of the reactive fuel jet and coordinate system used in the  derivation of the mathematical model. }
\label{fig:3} 
\end{figure}

In this section, we present a derivation of  an elementary model for the propagation of an advancing autoignition front in a fuel/oxidizer co-flow jet injected into 
water at high pressure.
Similar to several of our previous models of autoignition  \cite{Jet1,Jet2,Jet4},  the present model is based on the combinations of Burke-Schumann theory of the diffusion flames and Semenov-Frank-Kamenetskii
theory of thermal explosion. In what follows, we assume that a co-flow fuel inner jet  (for example consisting of a mixture of ethanol and water)  and  oxidizing outer jet of air is injected into ambient consisting of  water in supercritical state. The model derived in this work attempts to capture some basic qualitative effects of the dynamics 
of the ignition front. The derivation requires making numerous simplifying assumptions which makes the resulting  model mathematically tractable. Our assumptions are  as follows:

\begin{itemize}
\item The fuel part of the co-flow jet assumes  a fixed cylindrical shape with non-necessarily round cross section $\Sigma$. Moreover, since the characteristic size of the
cross-section is much smaller than the heights of the jet, we regard the jet as infinite in the direction of the injection (see Figure \ref{fig:3}).

\item The velocity of the jet $u$ is regarded as a constant.

\item The initial temperature of the jet $T_0$ is well below the effective ignition temperature.

\item The reaction takes place only on the surface  separating fuel and oxidizer parts of the co-flow jet (reaction surface).

\item The reaction on the reaction surface  is relatively weak and does not impact hydrodynamic structure of the jet.

\item As thermal conductivity and the heat loss within the oxidizing outer part of the jet are substantially smaller than the one of the fuel/water part    of the jet, the entire heat flux generated by the chemical reaction of the surface of the jet is directed
toward the fuel  part.

\item Both the fuel and the oxidizer on the surface of the jet are in excess,  and their consumption by a weak chemical reaction is negligible. Hence,
concentrations of both fuel and the oxidizer  on the reaction surface will be regarded as constant.

\item As a consequence of the previous assumption, the  intensity of the reaction on the surface of the jet depends exclusively on temperature.

\item In view of three previous assumptions, the dynamics of the ignition front within co-flow jet is determined by temperature distribution in the fuel part of the
jet and hence the oxidizing part of the jet will not be considered.

\item The principal components that govern the temperature distribution within the fuel part of the  jet are: the advection in the vertical direction, thermal diffusion and the heat loss which are
balanced by the heat production on the boundary of the fuel part of the jet.

\item There are two stable steady states of temperature within the cross section independent of the position along the jet. The first one corresponds to a nonreactive state 
and the second one to a reactive state. (The formal definition of the stability is presented below).
\end{itemize}
Under these assumptions, the conservation of energy equation within the fuel part of the jet reads:
\begin{eqnarray}\label{eq:m1}
c_p\rho \left(\frac{ \partial}{\partial t} T-u\frac{\partial }{\partial z} T\right)=\kappa \tilde \nabla^2 T-G(T).
\end{eqnarray}
Here, $T$ is the temperature, $c_p$ and $\rho$ are the specific heat and density of the fuel/water mixture respectively, $u$ is the velocity of the jet regarded as constant,
$\kappa$ is thermal conductivity,
$t$ is time,  $\tilde \nabla^2=\frac{\partial^2}{\partial x^2}+\frac{\partial^2}{\partial y^2}+\frac{\partial^2}{\partial z^2}$ is the diffusion (Laplace) operator, $(x,y,z)$ are spatial coordinates with coordinate $z$  aligned with the center line of the jet and directed agains the flow direction, and $G(T)$ is a heat loss function.

The equation \eqref{eq:m1} holds in the cylindrical domain $(x,y,z)\in  \Sigma \times \mathbb{R}$ occupied by the fuel part of the jet and for time $t>0$. 
As stated in the assumptions above, the heat flux toward the fuel part of  jet  is specified as follows: 
\begin{eqnarray}\label{eq:m3}
\kappa \tilde \nabla T\cdot {\bf n}=F(T),
\end{eqnarray}
where $\tilde \nabla$ is the gradient,  ${\bf n}$ is an outward normal to the reaction sufrace, and $F(T)$ is a function prescribing reaction.

In view that we are interested in the  traveling front solutions of \eqref{eq:m1}, \eqref{eq:m3}, we introduce a traveling front ansatz:
\begin{eqnarray}\label{eq:m4}
T(x,y,z,t):=\tilde T(x,y,\bar z), \quad \bar z=z-vt,
\end{eqnarray}
where $v$ is an  a-priori unknown velocity of the ignition front in a laboratory frame. Substitution of \eqref{eq:m4} into \eqref{eq:m1} gives:
\begin{eqnarray}\label{eq:m5}
-c_p\rho (u+v) \frac{\partial}{\partial \bar z}  \tilde T=\kappa \tilde \nabla^2 \tilde T-G(\tilde T).
\end{eqnarray}

For further studies, it is convenient to introduce non-dimensional quantities. Namely, non-dimensional temperature $\theta$ and non-dimensional coordinates $(\xi,\eta,\zeta)$
defined as follows:
\begin{eqnarray}\label{eq:m6}
\theta=\frac{\tilde T-T_0}{T_0}, \quad \xi=\frac{x}{R}, \quad \eta=\frac{y}{R}, \quad \zeta=\frac{\bar z}{R},
\end{eqnarray}
here $T_0$ is the initial temperature of the jet, $R$ is the characteristic  size of the cross section which is taken as the smallest radius of the disc into which the cross section of the jet can be  inscribed.
With this normalization, we denote the non-dimensional cross section of the jet as $\Omega:=R^{-1} \Sigma$ which has characteristic size of unity.
 
In terms of these non-dimensional quantities equations, \eqref{eq:m5} and \eqref{eq:m3} take the following form:
\begin{eqnarray}\label{eq:m7}
-c\frac{\partial}{\partial \zeta} \theta =\nabla^2 \theta-h g(\theta), \quad (\xi,\eta,\zeta)\in \Omega \times   \mathbb{R},
\end{eqnarray}
 \begin{eqnarray}\label{eq:m8}
 \nabla \theta \cdot  {\bf \nu }=q f(\theta) \quad ( \xi, \eta, \zeta) \in \partial \Omega  \times \mathbb{R}.
 \end{eqnarray}
 Here, $\nabla^2 \theta=\left (\frac{\partial^2}{\partial \xi^2}+\frac{\partial^2}{\partial \eta^2}+\frac{\partial^2}{\partial \zeta^2}\right) \theta$,~
 $\nabla \theta\cdot \nu=\nu_{\xi} \frac{\partial \theta }{\partial \xi}+\nu_{\eta} \frac{\partial \theta }{\partial \eta}$, $\nu=(\nu_{\xi},\nu_{\eta})$ is the
 unit outward normal to the boundary of $\Omega$ denoted as $\partial \Omega$, $g(\theta)$ and $f(\theta)$ are non-dimensional heat loss and reaction functions,
 $h$ and $q$ are non-dimensional heat loss  and reaction intensity parameters and 
\begin{eqnarray}\label{eq:m9}
c=\frac{c_p\rho (u+v) R}{\kappa},
\end{eqnarray}
is non-dimensional speed of propagation of the ignition  front relative to gas flow.

In what follows, we assume that the heat loss function $g(\theta)$ is a sufficiently regular convex increasing function satisfying $g(0)=0,$ $g^{\prime}(0)>0.$
These assumptions are general enough to incorporate most types of heat loss mechanisms, in particular, radiative heat loss which is
most relevant to the current study. In this case $g$ is given by:
\begin{eqnarray}\label{eq:m10}
g(\theta)=(1+\theta)^4-1, \quad h=\frac{4 \sigma K_p P T_0^3 R^2}{\kappa},
\end{eqnarray}
where $\sigma$ is the Stefan-Boltzmann constant, $K_p$ is the Planck mean absorption coefficient and $P$ is  ambient pressure.

\begin{figure}[h!]
\hspace{-.5cm}
%\noindent %
\begin{minipage}[c]{0.4\textwidth}%
%\begin{center}
%\hspace{-2.5cm}
\includegraphics[width=3.in]{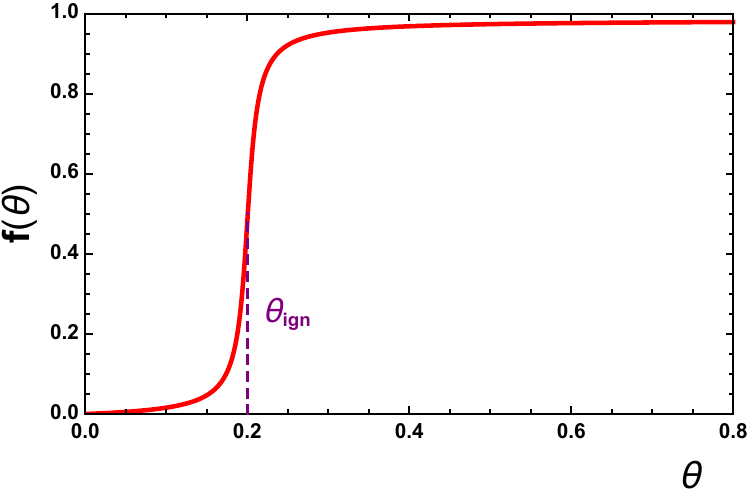}
%\vspace{-1in}
%\llap{\raisebox{1.15cm}{
% \includegraphics[width=1.1in]{333}
%}}
%\hspace{3.977cm}
%\llap{\raisebox{4.cm}{
% \includegraphics[width=2.2in]{333a}
%}}
%\par\end{center}%
\end{minipage}\hspace{1.2cm} %
\hspace{.5cm}
\begin{minipage}[c]{0.4\textwidth}%
 \includegraphics[width=3.in]{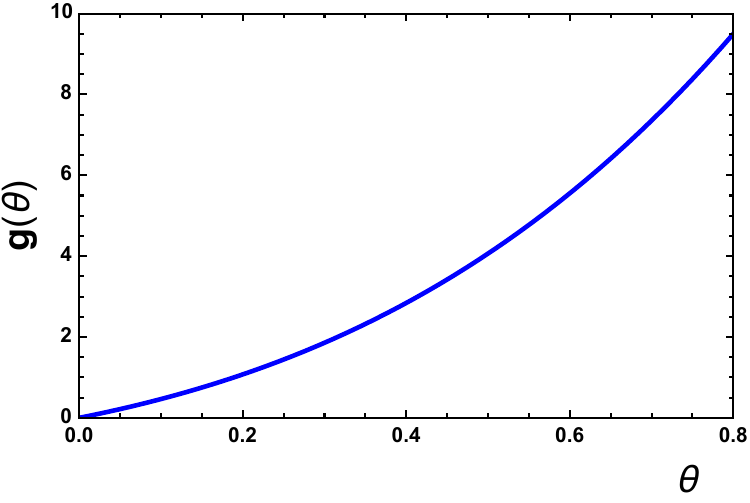} %
\end{minipage}
\caption{Characteristic profiles of reaction intensity and heat loss functions. The heat loss function $g$ depicted on the right panel of this plot is given by 
\eqref{eq:m10} and the reaction intensity function $f$ depicted on the left panel of this plot is given by \eqref{eq:m10a} with $\theta_{ign}=0.2$ 
and $d=10^{-2}$.}
\noindent \label{fig:4} 
\end{figure}

The  reaction function $f(\theta)$ is assumed to be an increasing function of  sigmoid shape. Specifically, we assume that
$f(0)=0,$ and slowly increases for temperatures smaller and larger than the effective ignition temperatures while experiencing rapid growth 
in the vicinity of the effective ignition temperature. We also normalize this function is such a way that $\lim_{s\to\infty} f(s)=1$. This assumption allows
 the incorporation of typical reaction functions such as  Arrhenius  and ignition types.
The non-dimensional  reaction intensity parameter is defined as:
\begin{eqnarray}\label{eq:m11}
 q=\frac{Q C_o C_f R}{\kappa T_0},
\end{eqnarray}
where $Q$ is an effective reaction intensity, $C_f$ and $C_o$ concentrations of the fuel and the oxidizer on the surface of the jet are regarded as prescribed constants.
In the numerical  simulations presented in the proceeding sections, we will use the following representation of the reaction term:
\begin{eqnarray}\label{eq:m10a}
f(\theta)=\frac{1}{\pi}\left(\arctan\left(\frac{\theta-\theta_i}{d}\right)+\arctan\left(\frac{\theta_i}{d}\right)\right),
\end{eqnarray}
where $\theta_i$ is an effective ignition temperature and $d$ is a sufficiently small positive number ($d=10^{-2}$ in numerical simulations presented in proceeding sections)  that ensures a rapid  change of the reaction function near the ignition temperature. 
The representative plots of the functions $f$ and $g$ are shown in  Figure \ref{fig:4}.

Problem \eqref{eq:m7}, \eqref{eq:m8} should be complemented with boundary like conditions for far fields of temperature
 (far behind and far ahead of the ignition front),
that is, conditions when $\zeta\to\pm\infty$. Hence, we require that
\begin{eqnarray}\label{eq:m12}
\theta(\xi,\eta,\zeta) \to \Theta^\pm(\xi,\eta)\quad  \mbox{in} \quad \Omega \quad \mbox{as}  \quad \zeta \to \pm \infty.
\end{eqnarray}
Far ahead of the front, the temperature is expected to be equal to the initial one. Thus,
\begin{eqnarray}\label{eq:m13}
\Theta^+=0.
\end{eqnarray}
Far behind the front, we require that the jet is in a reactive state. Therefore, $\Theta^-$ is the largest positive solution of the following problem:
\begin{eqnarray}\label{eq:m14}
\left\{
\begin{array}{ll}
\Delta  \Theta^- = h g(\Theta^-), & (\xi,\eta)\in \Omega,\\
\nabla \Theta^-\cdot \nu = q f(\Theta^-), & (\xi,\eta)\in \partial \Omega,
\end{array}
\right.
\end{eqnarray}
where $\Delta=\frac{\partial^2}{\partial \xi^2}+\frac{\partial^2}{\partial \eta^2}$ is the diffusion operator within the cross-section of the jet.

We also require that both $\Theta^{\pm}$ are stable solutions. Stability, in this context,  means that the principal eigenvalue of the linearization of the corresponding problems are positive.
More precisely, we assume that the principal (smallest)  eigenvalues  $\mu_{\pm}$, that is smallest value of parameters $\mu_{\pm}$ for which problem
\begin{eqnarray}\label{eq:m15}
\left\{
\begin{array}{ll}
-\Delta \phi^{\pm} +h g^{\prime}(\Theta^{\pm})=\mu_{\pm} \phi^{\pm}, & (\xi,\eta)\in \Omega,\\
\nabla \phi^{\pm} \cdot \nu = q f^{\prime}(\Theta^{\pm}), & (\xi,\eta)\in \partial \Omega,
\end{array}
\right.
\end{eqnarray}
admits non-trivial solutions,  are strictly positive ($\mu_{\pm}>0$).  Here and below, $f^{\prime}(s)=df(s)/ds, ~ g^{\prime}(s)=dg(s)/ds$. These conditions ensure that the ignition front connects two stable states $\Theta^{\pm}$.
The presence of  stable steady states $\Theta^{\pm}$ depends quite sensitively on the specific values of the parameters $h$ and $q$. This  will be discussed in the next section.

Under the assumptions presented in the beginning of this section, the study of  the propagation of traveling ignition fronts  reduces to the analysis of
problem \eqref{eq:m7}, \eqref{eq:m8}, \eqref{eq:m12}. From a mathematical perspective, this is a classical problem in the  theory of reaction-diffusion equations which is
  finding traveling front solutions in a cylindrical domain, in other words,  finding  a pair(s)  $(c,\theta)$ that verifies  \eqref{eq:m7}, \eqref{eq:m8}, \eqref{eq:m12}.
  While poofs of the  existence and uniqueness of traveling front solutions for problem \eqref{eq:m7}, \eqref{eq:m8}, \eqref{eq:m12} as stated are not available in the existing literature,
  it can be obtained by minor modifications of the theories presented in \cite{Vega,Heinze} (see also \cite{kyed,volpert2,muratov}) and will be omitted here. 
  Using these approaches, one can show that there exists a unique pair $(c^*,\theta)$ that verify  \eqref{eq:m7}, \eqref{eq:m8} and connects minimal 
  and maximal stable steady states $(\Theta^{\pm})$ far ahead and behind the front provided the latter exists and is stable. We note the  uniqueness holds up to translations
   along the central line of the jet due to translational invariance of the problem with respect to vertical coordinate $\zeta$. In what follows, to fix the translations we require that the temperature at the origin is
  equal to the effective ignition temperature.
  Our goal here will be to analyze qualitative properties of 
  solutions of \eqref{eq:m7}, \eqref{eq:m8}, \eqref{eq:m12}.  
  We note that the existence of the solution for \eqref{eq:m7}, \eqref{eq:m8}, \eqref{eq:m12} does not guarantee successful autoignition. Indeed, to ensure that the ignition
  of the jet takes place, the advancing front has to propagate towards the injection inlet with a positive, in  a laboratory frame,  speed. Hence, the autoignition condition is as follows:
  \begin{eqnarray}\label{eq:m16}
  c^{\sharp}=c^*-c^\dagger>0.
  \end{eqnarray}
  That is, the velocity of the front in a laboratory frame $c^*$ (the velocity of traveling front obtained as the solution of  problem \eqref{eq:m7}, \eqref{eq:m8}, \eqref{eq:m12}) has to exceed the velocity of the flow ${c^\dagger}=\frac{c_p\rho u R}{\kappa}.$ In terms of dimensional quantities, this condition
  reads:
  \begin{eqnarray}\label{eq:m17}
  c^* \frac{\kappa}{c_p\rho R}>u.
  \end{eqnarray}
  For fixed reaction, heat loss mechanisms and flow speed, the non-dimensional velocity of propagation relative to the flow $c^*$ fully determines whether autoignition of the jet takes place.
  The velocity $c^*$ depends, in a rather nontrivial way, on two parameters of the problem: non-dimensional reaction intensity $q$ and non-dimensional heat loss parameter $h$. 
  In view that physicochemical parameters of the experimental system of interest can vary substantially, 
 it is essential to consider possible scenarios of propagation of the ignition fronts when  $h$ and $q$  vary in a wide range. 
 In the following section, we will analyze feasible parametric regimes of propagations when $h$ and $q$ change and discuss some physical implications
 of the analysis.

 \section{Analysis of the model}\label{s:analysis}

 In this section, we discuss feasible parametric  regimes of propagation of the advancing ignition fronts.  As was noted in the previous section, the sharp autoignition condition
 \eqref{eq:m16} requires that the non-dimensional velocity of the ignition front relative to the flow $c^*$ exceeds non-dimensional flow velocity $c^\dagger$.
As $c^{\dagger}$ is viewed as given and can be defined from the experimental set up, $c^*$ is determined from the solution of  \eqref{eq:m7}, \eqref{eq:m8}, \eqref{eq:m12}. The solution of this problem is fully determined by two parameters: non-dimensional reaction intensity $q$ and the heat loss parameter $h$.
In what follows, we will discuss the  range of parameters $q$ and $h$ for which  solutions of problem \eqref{eq:m7}, \eqref{eq:m8}, \eqref{eq:m12} exist and are physically meaningful
and discuss solutions of this problem in these parametric regimes.

\begin{figure}[h]
\noindent \begin{raggedright}
\hspace*{0.5cm}%
\begin{minipage}[c]{0.9\textwidth}%
\includegraphics[width=6in]{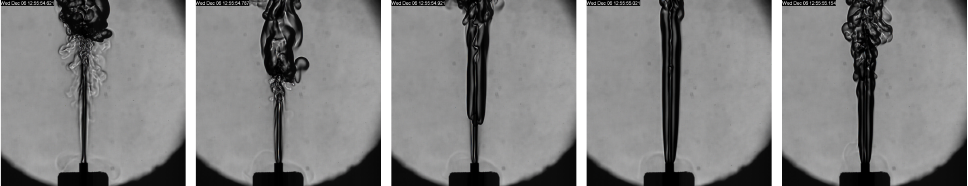} %
\end{minipage}\hspace*{0.3cm} %
\vspace*{\bigskipamount}
\par\end{raggedright}
\noindent \begin{raggedright}
\hspace*{0.5cm}%
\begin{minipage}[c]{0.9\textwidth}%
\includegraphics[width=6.in]{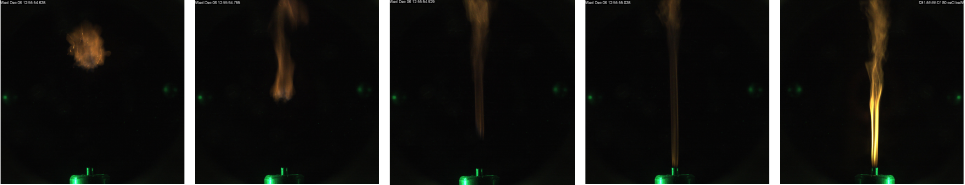} %
\end{minipage}\hspace*{0.3cm} %
\par\end{raggedright}
\noindent \raggedright{}\caption{{Series of consequent images showing the realization of the reaction dominated regime. In this regime spontaneous formation of the ignition kernel 
 instantaneously ignites the flame. Shadowgraphic images on the upper panel show  variations  in temperature field, whereas  images on the lower panel 
 show the visible flame.}}
\label{fig:5} 
\end{figure}

Recall that parameter $q$ measures the strength of the reaction relative to 
 thermal diffusivity. If the reaction strength is much larger than the thermal diffusivity (that is $q\gg 1$),  the formation of the ignition kernel results in the instantaneous ignition of the flame
 (see Figure \ref{fig:5}).
 This {\it reaction dominated} regime  is irrelevant to the current study as no ignition front appears in this case. The analysis of autoignition in this regime reduces to identifying
 conditions for the formation of the ignition kernel and can be studied in the  framework of our model proposed in \cite{Jet1}.
Therefore, we need to consider two situations:  when $q$ is small and when $q$ is of the order of unity. The first regime
physically corresponds to the situation when diffusion dominates reaction and the second one when diffusion and reaction are comparable. Both of these regimes are of interest
and will be analyzed below. 
We start with the analysis of the first regime in which $0<q\ll 1.$  We will call this regime {\it  diffusion dominated} as in this regime
both reaction intensity $q$ and the heat loss parameter $h$ must be small and comparable to each other. Hence, thermal diffusion is the principle
mechanism driving the ignition front. Moreover, in this regime  the temperature within each fixed cross section of the reactive  jet
is close to a constant, and therefore problem  \eqref{eq:m7}, \eqref{eq:m8}, \eqref{eq:m12} can be further simplified and reduced to a one-dimensional
reaction-diffusion equation with bistable non-linearity that describes the evolution of an average over the  jet's cross-section temperature within ignition front.  Indeed, as shown in the Appendix in this regime  the  requirement of the existence of  a reactive state $\Theta^{-}$
is only met when the heat loss parameter $h$ is also small. In addition, it is shown that when both $q$ and $h$ are small the variations of $\Theta^{-}$ over the cross
section also needs to be small. Consequently, $\Theta^{-}$ is approximately constant. This immediately implies that the ignition front in this regime
approaches to  essentially constant states $\Theta^{\pm}$ as $\zeta\to \pm \infty,$ and hence, the traveling front is effectively one-dimensional.
These formal arguments can be made rigorous using the methods of \cite{Heinze, freidlin}.

To obtain an effective equation describing the propagation of the ignition front in this regime,  let us observe that the values of functions $\Theta^{-}$ and $\theta$ are nearly constant
over cross sections of the jet. Therefore, for the fixed position along the central line of the jet the values of $\Theta^{-}$ and $\theta$ are determined, in the first approximation,  by  their averages over cross section which coincide with the averages over the boundary of the cross sections.
Hence, setting $\bar \Theta^{-}$ and $\bar \theta(\zeta)$ to be averages of $\Theta^{-}(\xi,\eta)$ and $\theta(\xi,\eta, \zeta)$ over a cross section we have: 
\begin{eqnarray}\label{eq:r1}
\bar \Theta^{-} =\frac{1}{|\Omega|} \int_{\Omega} \Theta^{-} (\xi,\eta) dA \simeq \frac{1}{|\partial \Omega|} \int_{\partial \Omega}  \Theta^{-}(\xi,\eta)dS, 
\end{eqnarray}
\begin{eqnarray}\label{eq:r2}
\bar \theta (\zeta) =\frac{1}{|\Omega|} \int_{\Omega} \theta (\xi,\eta,\zeta) dA \simeq \frac{1}{|\partial \Omega|} \int_{\partial \Omega}  \theta(\xi,\eta,\zeta)dS, 
\end{eqnarray}
where $dA,$ $dS$ are  elements of an area and  of a boundary of a cross-section of the jet respectively,
$|\Omega|$ is the area of cross-section of the jet,  $|\partial \Omega|$ is the perimeter of the cross section of the jet. The sign $\simeq$ stands for asymptotically equivalent.
We note that if a cross-section is the unit disk   $|\Omega|=\pi$ and $|\partial \Omega|=2\pi.$

Now observe that
\begin{eqnarray}\label{eq:r3}
\int_{\Omega} \frac{\partial}{\partial \zeta} \theta(\xi,\eta,\zeta) d A= |\Omega|  \frac{\partial}{\partial \zeta} \bar \theta (\zeta),
\end{eqnarray}
and
\begin{eqnarray}\label{eq:r4}
&&\int_{\Omega} \nabla^{2}  \theta (\xi,\eta,\zeta) dA=\int_{\Omega}  \frac{\partial^2}{\partial \zeta^2} \theta(\xi,\eta,\zeta)dA+\int_{\Omega} \left( \frac{\partial^2}{\partial \xi^2}+
\frac{\partial^2}{\partial \eta ^2}\right)
 \theta(\xi,\eta,\zeta)dA=\nonumber\\
 && |\Omega|  \frac{\partial^2}{\partial \zeta^2} \bar \theta(\zeta)+\int_{\partial \Omega} \nabla \theta(\xi,\eta,\zeta) \cdot \nu dS=
 |\Omega|  \frac{\partial^2}{\partial \zeta^2} \bar \theta(\zeta)+q\int_{\partial \Omega} f(\theta(\xi,\eta,\zeta)) dS,
\end{eqnarray}
where the second equality follows from the Green's identity, and the third equality follows from the boundary condition in \eqref{eq:m7}.
Moreover, by \eqref{eq:r2} we have:
\begin{eqnarray}\label{eq:r5}
\int_{\Omega} g(\theta(\xi,\eta,\zeta))dA\simeq |\Omega| g(\bar \theta(\zeta)), \quad \int_{\partial \Omega} f(\theta(\xi,\eta,\zeta))dS\simeq |\partial \Omega| f(\bar \theta(\zeta)).
\end{eqnarray}
Hence  integration of  \eqref{eq:m7} over a cross section,   taking into account \eqref{eq:r3}, \eqref{eq:r4}, \eqref{eq:r5},  yields the following equation for the average temperature:
\begin{eqnarray}\label{eq:an11}
 - c \frac{d}{d \zeta} \bar \theta =\frac{d^2}{d  \zeta^2} \bar \theta+ q \frac{|\partial \Omega|}{|\Omega|}f( \bar \theta)- h g(\bar \theta), \quad  \zeta \in \mathbb{R}.
 \end{eqnarray} 
Moreover, integrating equation \eqref{eq:m14} over the cross section of the jet and taking into account Green's identity and the boundary condition in \eqref{eq:m14} we have:
\begin{eqnarray}\label{eq:r6}
h \int_{\Omega} g(\Theta^{-}(\xi,\eta)) dA=\int_{\Omega} \Delta \Theta^{-}(\xi,\eta) dA =\int_{\partial \Omega}  \nabla \Theta^{-}(\xi,\eta) \cdot \nu dS=
q\int_{\partial \Omega}  f(\Theta^{-}(\xi,\eta)) dS.
\end{eqnarray}
Since by \eqref{eq:r1}
\begin{eqnarray}\label{eq:r7}
 \int_{\Omega} g(\Theta^{-}(\xi,\eta) dA\simeq |\Omega| g(\bar \Theta^{-}), \quad \int_{\partial \Omega}  f(\Theta^{-}(\xi,\eta)) dS\simeq |\partial \Omega| f(\bar \Theta^{-}),
\end{eqnarray}
we obtain that:
\begin{eqnarray}\label{eq:r8}
h |\Omega| g(\bar \Theta^{-})= q|\partial \Omega| f(\bar \Theta^{-}).
\end{eqnarray}
Hence in this regime, the boundary value problem \eqref{eq:m14} reduces to the transcendental equation \eqref{eq:r8}.

\begin{figure}[h]
%\centering \includegraphics[width=4in]{fig6} 
%\hspace{1.5cm}
%\noindent %
\begin{minipage}[c]{0.45\textwidth}%
%\vspace{.8cm}
%\begin{center}
%\hspace{-2.5cm}
\includegraphics[width=3.in]{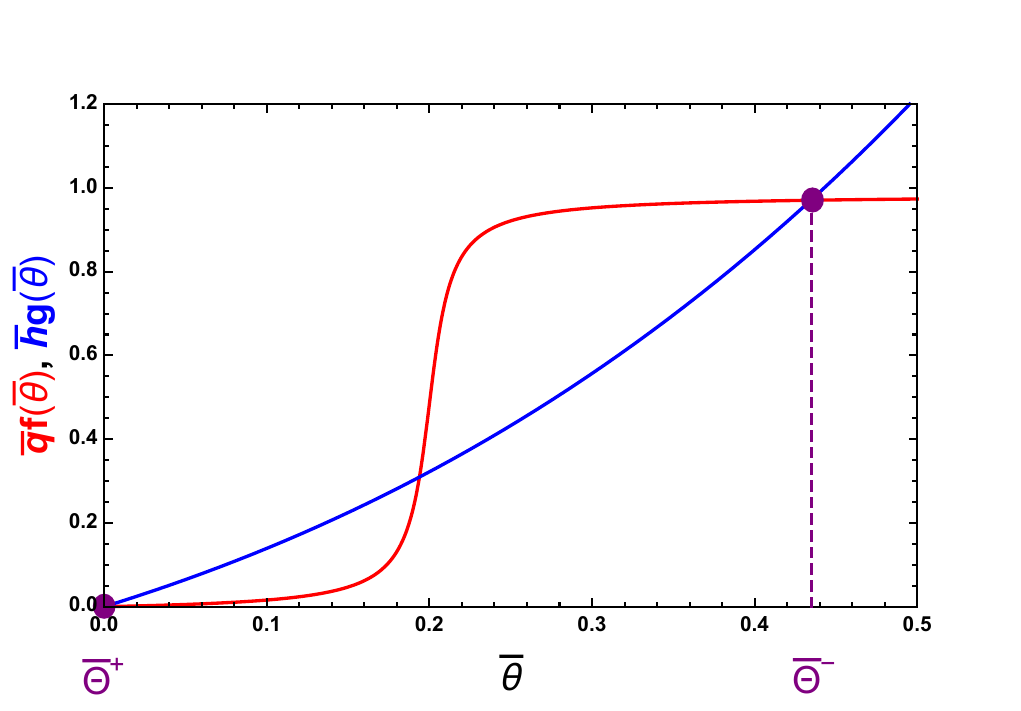}
%\vspace{-1in}
%\llap{\raisebox{1.15cm}{
% \includegraphics[width=1.1in]{333}
%}}
%\hspace{3.977cm}
%\llap{\raisebox{4.cm}{
% \includegraphics[width=2.2in]{333a}
%}}
%\par\end{center}%
\end{minipage}\hspace{1.2cm} %
\hspace{.5cm}
\begin{minipage}[c]{0.45\textwidth}%
 \includegraphics[width=3.in]{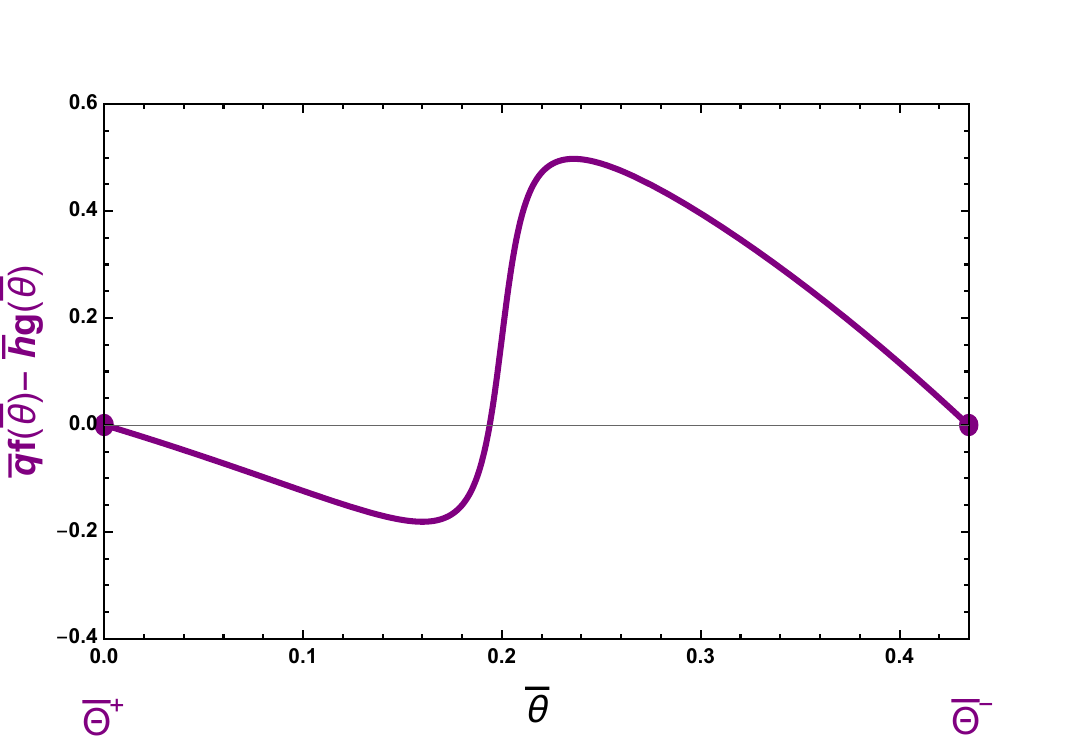} %
\end{minipage}
\caption{Left panel: limiting values  for  far field of the ignition front $\bar \Theta^{\pm}$ defined as an intersection of the scaled reaction function ($\bar q f(\bar \theta)$) and 
scaled heat loss functions ($\bar h g(\bar \theta)$) with $\bar \Theta^{+}=0$ and $\bar \Theta^{-}\approx 0.435$ being the largest intersection. 
Right panel: the overall nonlinear term $\bar q f(\theta)-\bar h g(\bar \theta)$.
In both figures the reaction function
is given by \eqref{eq:m10a} with $\theta_{ign}=0.2$ and $d=10^{-2}$ and heat loss function given by \eqref{eq:m10}. The scaled heat loss parameter $\bar h=0.3$ and scaled
reaction intensity $\bar q=1$.  }
\label{fig:6} 
\end{figure}

Since both $q$ and $h$ are small in the diffusion dominated regime,  it is convenient  to introduce the  following rescaling of the parameters of the problem and coordinate $\zeta$,
\begin{eqnarray}\label{eq:an12}
 h=\eps \bar h, \quad q=\eps \frac{|\Omega|}{|\partial \Omega|} \bar q \quad \zeta=\frac{\bar \zeta}{\sqrt{\eps}} , \quad  c=\sqrt{\eps} \bar c,
 \end{eqnarray}
 where $0<\eps\ll 1$ is a small parameter.  In this rescaling \eqref{eq:an11} reads:
 \begin{eqnarray}\label{eq:an13}
 -\bar c \frac{d}{d\bar \zeta} \bar \theta =\frac{d^2}{d \bar \zeta^2} \bar \theta+\bar q f(\bar \theta)-\bar h g(\bar \theta), \quad \bar \zeta \in \mathbb{R}
 \end{eqnarray} 
 The far field conditions take the form:
 \begin{eqnarray}\label{eq:an14}
  \bar \theta(\zeta)\to \bar \Theta^{\pm} \quad \mbox{as}\quad \bar \zeta \to \pm \infty.
 \end{eqnarray}
 with $\bar \Theta^+=0$ and $\bar \Theta^-$ being the largest solution of transcendental equation
 \begin{eqnarray}\label{eq:an15}
\bar q  f(\bar \Theta^-)=\bar h g(\bar \Theta^-).
 \end{eqnarray}
For fixed $f$ and $g$ depending on particular choices of $\bar h$ and $\bar q$,  this equation gives either two, one or no positive solutions. 
The only case which is of interest for the current study is the case when \eqref{eq:an15} has two positive solutions since this is the
only case when both $\Theta^{\pm}$ exist and are stable.
The stability condition \eqref{eq:m15} in this case reduces to $\bar q f^{\prime}(\bar \Theta^{\pm})<\bar h g^{\prime}(\bar \Theta^{\pm})$.
This situation is realized if the ratio of $\bar h$ and $\bar q$ does not exceed some critical value of order one. The typical situation is depicted in
 Figure \ref{fig:6}. We also note  that the ratio $\bar h$ and $\bar q$ can't also be too small. Indeed, if $\bar h/ \bar q$ is small the value of $\Theta^{-}$
 will be large which  physically contradicts  the assumption that the relative elevation of the temperature in the reactive state is of order one.
 This, in particular, implies that $\bar q$ and $\bar h$ are  of the same order and hence $q$ and $h$ are of the same order.

 \begin{figure}[h!]
\centering \includegraphics[width=4in]{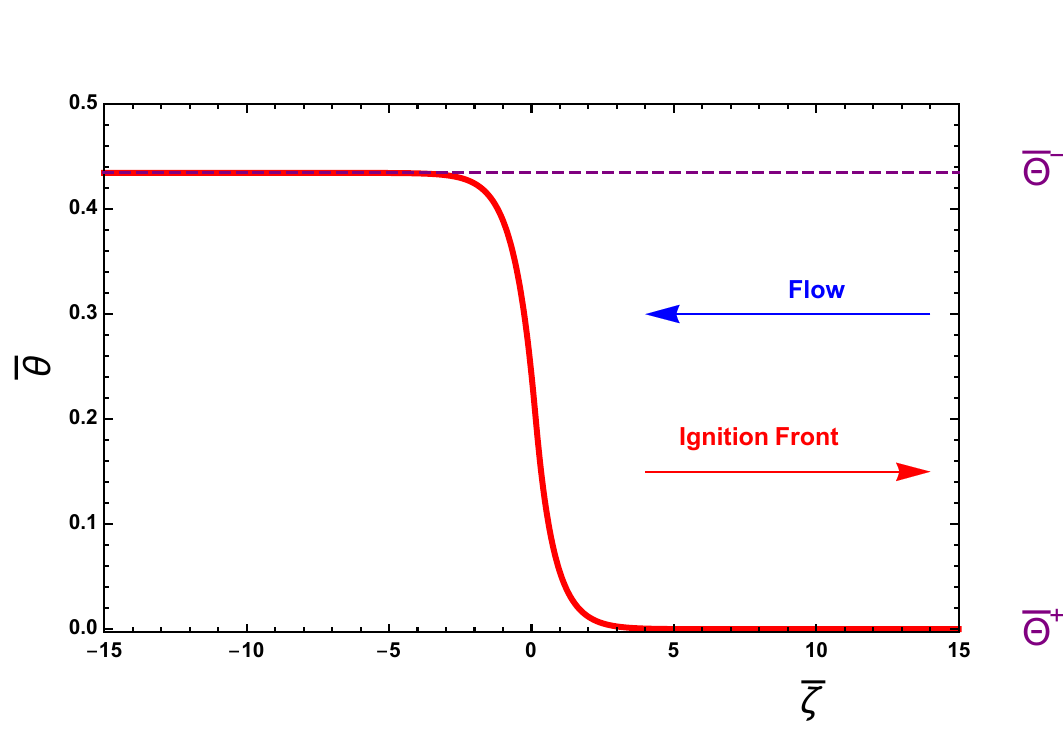} 
\caption{Traveling front profile $\bar \theta$ for problem with $\bar q=1$, $\bar h=0.3$ and $g$ given by \eqref{eq:m10} and $f$ given by \eqref{eq:m10a}
with $\theta_{ign}=0.2$. The velocity of propagation $\bar c^*\approx 0.77$ and reactive state $\bar \Theta^-\approx0.435$
 }
\label{fig:7} 
\end{figure}

Therefore, in the diffusion  dominated regime, problem  \eqref{eq:m7}, \eqref{eq:m8}, \eqref{eq:m12} effectively reduces to a one dimensional problem
\eqref{eq:an13}, \eqref{eq:an14} with  parameters $\bar q, \bar h$  of order one.
 This is a  classical reaction-diffusion problem with bistable non-linearity which was studied in detail in mathematical literature.
 In particular, it is well known that problem \eqref{eq:an13}, \eqref{eq:an14} admits a unique solution $(\bar c^*,\bar \theta)$  which is monotone
 and nonlinearly  stable \cite{Fife}. Moreover, the sign of the propagation velocity is fully determined by the integral of the nonlinear term. Specifically, 
 \begin{eqnarray}
 {\rm sign} (\bar c^*)= {\rm sign} \left (\int_{\Theta^+}^{\Theta^-} \left(\bar q f(s)-\bar h g(s)\right)ds\right).
 \end{eqnarray} 
 Thus, in view that  the velocity $\bar c^*$ for an advancing ignition front must be positive this  necessary condition requires that $\int_{\Theta^+}^{\Theta^-} \left(\bar q f(s)-\bar h g(s)\right)ds>0$. We note that for fixed values of parameters and nonlinearities $f,g$ problem \eqref{eq:an13}, \eqref{eq:an14}  can be easily solved 
 numerically (for example by shooting method). Figure \ref{fig:7} depicts a traveling front solution of problem  \eqref{eq:an13}, \eqref{eq:an14} with
 $g$ given by \eqref{eq:m10} and  $f$ given by \eqref{eq:m10a}, $\bar q=1$ and $\bar h=0.3,$ 
 $\theta_{ign}=0.2$ and $d=10^{-2}$. The overall nonlinear term $\bar q f(\bar \theta)-\bar h g(\bar \theta)$ for this set of parameters is depicted on the second panel of Figure \ref{fig:6}. In this case, the average temperature in the reactive state is $\bar \Theta^-\approx0.435$ (see first panel of Figure \ref{fig:6}) and the
scaled velocity of propagation  is $\bar c^*\approx 0.77$.
Let us also note that in the diffusion dominated regime the presence of an advancing ignition front is realized for the following scaling of non-dimensional parameters of the problem
  $h\sim \eps,$  $q\sim \eps$,   $c^* \sim \sqrt{\eps}$ , $c^{\dagger} \sim \sqrt{\eps}$ with $0<\eps\ll 1$.
 Consequently, the velocity of the jet in a laboratory frame for successful propagation of the ignition front has to be of the
 same order as the velocity of the ignition front, that is  $c^{\sharp}\sim \sqrt{\eps}.$  Hence, $c^{\sharp}$ scales as a square root of thermal conductivity, and
 thus, the diffusion within the jet is a main mechanism driving the ignition front as claimed in the beginning of the discussion of this regime.
 We also note that the behavior of the ignition front is very sensitive to even small changes of the parameters. Indeed, in view that $c^{\sharp} \ll 1$ even 
 small changes in parameters of the problem can change $c^{\sharp}$ from being positive to being negative. Hence in the diffusion dominated regime,
 the advancing ignition front is on the verge of extinction. This regime was observed experimentally in \cite{exp}, and the typical structure of the 
 ignition front in this regime is depicted in Figure \ref{fig:2}. Finally, let us mention that for a given cross-section of the jet the effective reaction intensity
 parameter increases with the increase of its perimeter as evident from \eqref{eq:an12}. This in particular suggests that jets with non-round cross sections
 are easier to ignite than the ones with cross sections being a disk.

  \begin{figure}[h]
\centering \includegraphics[width=6in]{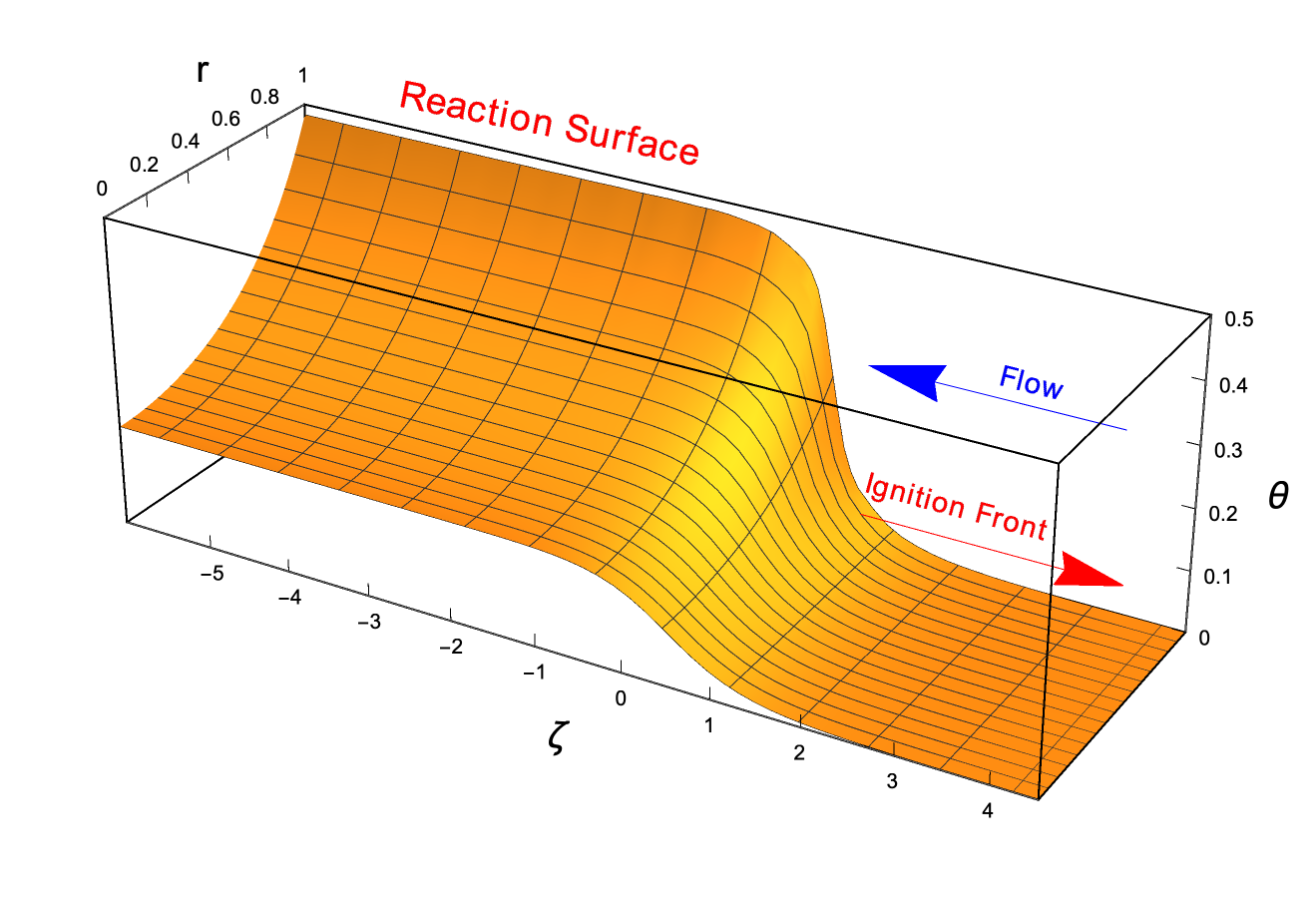} 
\caption{Numerical solution of problem \eqref{eq:m7}, \eqref{eq:m8}, \eqref{eq:m12}  representing the temperature distribution within an advancing ignition front  in the cylindrical reactive jet with round cross section. Here  $r=\sqrt{\xi^2+\eta^2}$ is the radial coordinate in the cross section of the jet. The heat loss 
and reaction intensity functions used in this numerical simulation are given by \eqref{eq:m10} and by \eqref{eq:m10a} with $\theta_i=0.2$ and $d=10^{-2}$ respectively.
The heat loss and reaction intensity parameters are set as $h=q=1$. 
The non-dimensional propagation velocity  $c^*\approx0.27$. Red arrow indicates direction of propagation of the ignition front and blue arrow indicated direction of flow. }
\label{fig:8} 
\end{figure}

\begin{figure}[h]
\hspace{1.5cm}
%\noindent %
\begin{minipage}[c]{0.2\textwidth}%
\vspace{.8cm}
%\begin{center}
%\hspace{-2.5cm}
\includegraphics[width=1.1in]{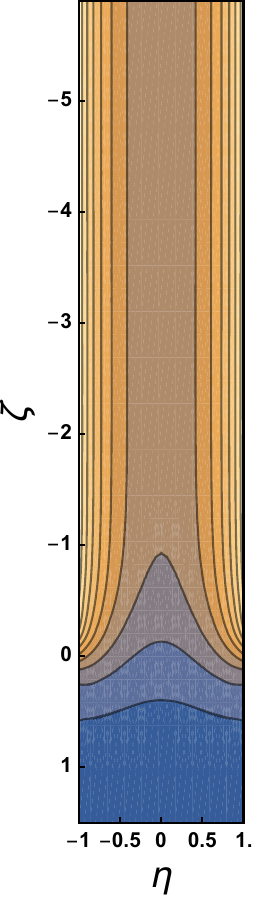}
\end{minipage}\hspace{1.2cm} %
\hspace{.5cm}
\begin{minipage}[c]{0.6\textwidth}%
 \includegraphics[width=3.5in]{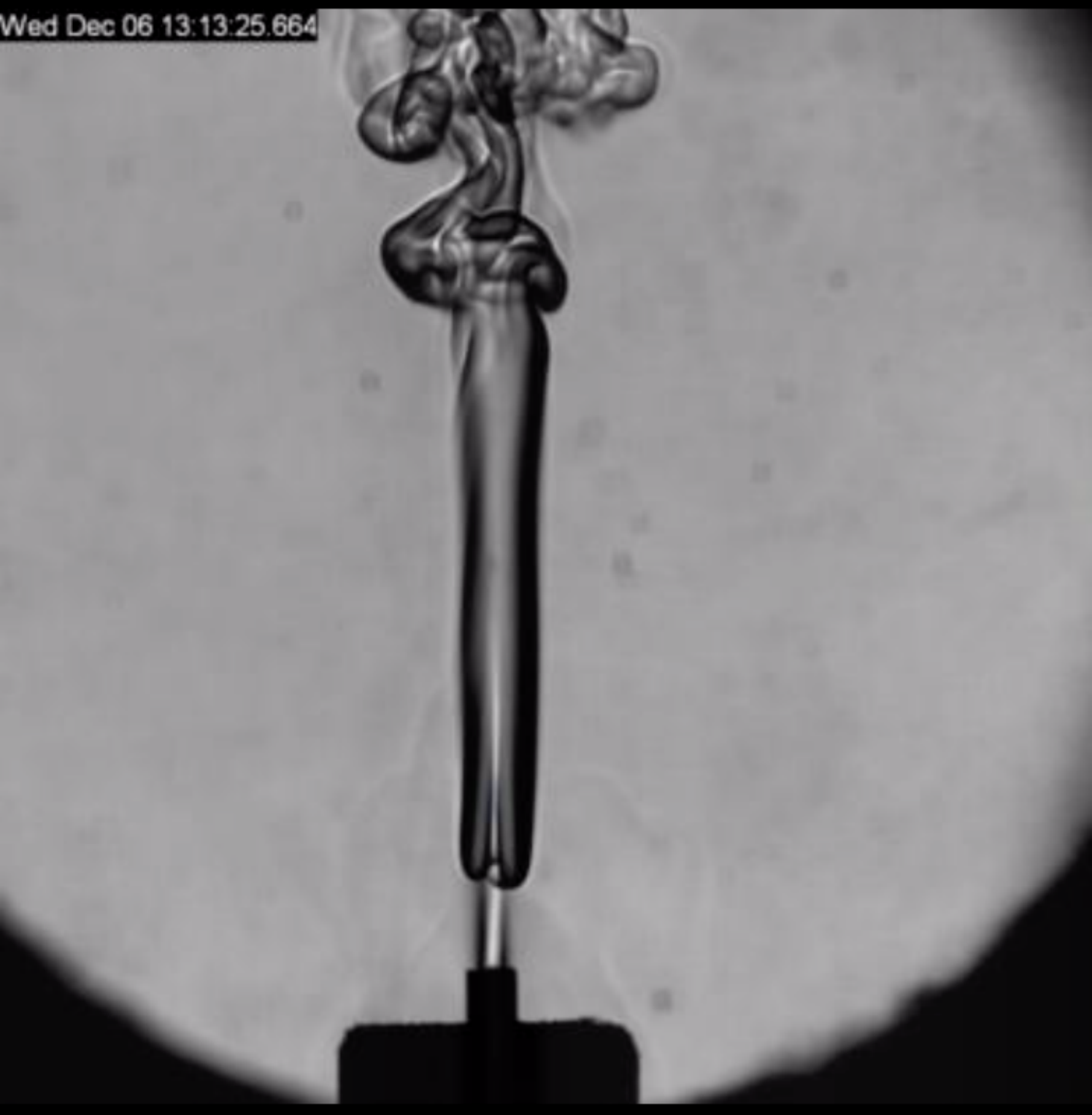} %
\end{minipage}\caption{{Isotherms of a slice $\xi=0$ of radial advancing ignition front depicted on Figure \ref{fig:8} and shadowgraph experimental image.  }}
\noindent \label{fig:9} 
\end{figure}

 We now turn to the situation when the reaction intensity is comparable with diffusion, that is, when  $q$ is  of order one. 
 We call this regime  {\it balanced} as in this situation diffusion, reaction and heat loss are comparable to each other.
 In contrast to the diffusion dominated regime, in the balanced regime the distribution of the temperature in each cross section is strongly uneven.
 Indeed, in the balanced regime the temperature within a fixed cross-section substantially rises from its smallest value in the central line of the jet to its largest value on the cross-section's boundary.
 We also note that in the balanced regime the velocity of the ignition front is of order one. This regime can be studied numerically.
 Figure \ref{fig:8} depicts an ignition front in the cylindrical  jet with round cross sections. In this numerical simulation,
 the reaction and heat loss function were taken as in \eqref{eq:m10} and \eqref{eq:m10a} with $\theta_{ign}=0.2$
 and $d=10^{-2}$. The values of heat loss and reaction intensity parameters were taken to be one, and
 the velocity of propagation of this ignition front turned out to be $c^*\approx0.27$.
This balanced regime was observed experimentally, and the numerical results are very much in line with the experiment in terms of predicting the shape of the ignition front
as evident from Figure \ref{fig:9}.

 \section{Concluding remarks}

 In this paper we proposed an elementary model  for  the propagation of  advancing ignition fronts in laminar reactive co-flow jets of fuel and oxidizer injected into an aqueous  environment at high pressure.  The model is derived under various simplifying assumptions which are quite in line with the existing experimental observations and utilizes a combination of Burke-Schumann  theory of diffusion flames and Semenov-Frank-Kamentskii theory of thermal explosion. The model consists  of 
 a single reaction-absorption-advection equation posed in an infinite cylinder with, not necessarily,  round cross sections and  a nonlinear condition on the cylinder's  boundary.
 The model derived in this work physically represents the conservation of energy within the  fuel part of the reactive jet. This equation  represents an interplay of diffusion, volumetric heat loss and advection within
 fuel part of the jet that are balanced by heat production on the reaction surface that separates fuel and oxidizer parts of the co-flow jet.
 The traveling front solution for this problem  corresponds to an advancing ignition front propagating against the flow and represents a temperature distribution within the fuel part of the co-flow jet that 
 connects reactive and nonreactive states far behind and far ahead of the front respectively.
 
  In a framework of the theory developed in this paper, we formulate a natural autoignition condition.
 This condition states that the velocity of the ignition front in a laboratory frame has to be positive. Furthermore, this condition allows the identification of  a range of physicochemical parameters of the problem 
 for which the propagation of the advancing ignition front takes place. The critical condition of successful autoignition can be restated as follows: the velocity of the ignition front
 relative to flow has to exceed the velocity of the flow. While the velocity of the flow is considered as given,  the velocity of the ignition front relative to the flow is obtained as a part of the solution
 of the boundary value problem  formulated in this paper. The velocity of the ignition front depends on two non-dimensional parameters which can be viewed as a ratio reaction strength
 and thermal conductivity and volumetric heat loss and thermal conductivity respectively. In experimental operating regimes, these parameters can vary in a wide range. Depending 
 on specific values of these parameters, several
 distinct modes of propagation of the ignition fronts take place.

The analysis of the model allowed the identification of three distinct regimes of autoignition all of which were observed experimentally in \cite{exp}. 
The first, reaction dominated regime,
is realized when the reaction intensity dominates diffusion. In this regime, the ignition process does not involve propagation of the ignition front
as ignition of the flame takes place right after the spontaneous formation of the ignition kernel. The second, diffusion dominated regime,  takes place
when both heat loss and reaction intensity are comparable to each other and much smaller than the characteristic diffusion. In this regime, the formation
 of the ignition front is followed by propagation of the ignition front with the temperature being nearly constant over the cross-sections of the front.
We show that   in the regime the model effectively reduces to a one-dimensional reaction-diffusion equation  with bistable non-linearity.
We also show that in the diffusion dominated regime the characteristic velocity of propagation scales as a square root of the diffusion and hence the ignition fonts is relatively slow. 
This observation implies that in the diffusion dominate regime even small variations in the parameters of the reactive system may change the direction of propagation
of the ignition front.  Hence in the diffusion dominated regime the  ignition front is on the verge of extinction. Finally, the third, balanced regime is realized when the reaction intensity 
and the heat loss are comparable to diffusion. In this case, the ignition front is represented by the genuine three dimensional structure which 
propagates much faster than the ignition front in the diffusion dominated regime. 

The model proposed in this paper provide a quantitative description of propagation of the advancing ignition fronts. The results of the analysis of the model given in this work
allow to estimation of a range of physical chemical and geometric parameters for which propagation of the ignition fronts takes place and, as we believe, contributes
to the physical understanding of this intriguing phenomena.
We also hope that the results presented in this paper will be useful for guiding future experimental studies of autoignition of hydrothermal flames.

\bigskip

\noindent {\bf Acknowledgments.} The work of MCH and UGH was supported  by NASA Biological and Physical Sciences Division in the Science Mission Directorate.
The work of AM and PVG was supported in a part by US-Israel Binational Foundation Grant 2020005.
Part of this work was performed when PVG was visiting NASA Glenn Research Center employed  by a contract with HX5 as a researcher in Summer 2023.
 PVG would like to thank  Low-Gravity Exploration Technology Branch of NASA Glenn Research Center for hospitality and HX5 staff for making implementation of this project possible. 
PVG also would like to thank Fedor Nazarov  and Gregory I. Sivashinsky for multiple enlightening discussions.

\bigskip

\noindent {\bf Declaration of competing interest. } The authors declare that they have no competing interests.
\bigskip

\section{Appendix}

In this appendix we  show that,  in the diffusion dominated regime, the  requirement that the  reaction intensity $q$ is small implies that  the heat loss parameter 
  $h$ must also be small and comparable to $q$. Moreover, we show that the function $\Theta^{-}$, if exists,  must be nearly constant in this regime. 
 
 Let us first prove that the diffusion dominated regime is realized if and only
if  the heat loss parameter  $h$ is  small. Clearly, a necessary condition for the existence of the solution for  \eqref{eq:m7}, \eqref{eq:m8}, \eqref{eq:m12} 
is the existence of a stable
reactive and non-reactive steady states $\Theta^{\pm}$. While non-reactive state $\Theta^+=0$ exists for arbitrary values of  $q$ and $h,$ 
  the reactive state 
$\Theta^-$   given as a solution of  \eqref{eq:m14} 
 can only exist when $h< c_0 q$, where constant $c_0>0$ is  of order unity. 
 Indeed,  multiplying equation \eqref{eq:m14} by $\Theta^{-},$   integrating the result over $\Omega$  and using Green's identity we have:
 \begin{eqnarray}\label{eq:an1}
 \int_{\partial \Omega} \Theta^{-} (\nabla \Theta^{-} \cdot \nu) dS=\int_{\Omega} |\nabla \Theta^{-}|^2dA+h \int_{\Omega} g(\Theta^{-}) \Theta^{-}dA,
 \end{eqnarray}
  where $dS$ and $dA$ stand for elements of the boundary and area of the jet's cross-section respectively. Taking into account the boundary condition in \eqref{eq:m14}
  this equality reads:
 \begin{eqnarray}\label{eq:an2}
q \int_{\partial \Omega} \Theta^{-} f(\ \Theta^{-})dS=\int_{\Omega} |\nabla \Theta^{-}|^2dA+h \int_{\Omega} g(\Theta^{-}) \Theta^{-}dA.
 \end{eqnarray}
In view that $f(0)=0$  and $f(s)$ is bounded, we have that $f(s)< c_1 s$ for all $s>0$ with some constant $c_1>0$ that depends only on $f.$ Moreover, since $g(0)=0, ~g^{\prime}(0)>0$ and $g(s)$ is increasing and convex for $s>0$ we have $g(s)\ge g^{\prime}(0)s$ for all $s\ge 0.$ These observations and \eqref{eq:an2} imply:
\begin{eqnarray}\label{eq:an3}
q c_1\int_{\partial \Omega} (\Theta^{-})^2dS\ge \int_{\Omega} |\nabla \Theta^{-}|^2dA+h g^{\prime}(0)  \int_{\Omega} (\Theta^{-})^2dA. 
\end{eqnarray}
Next  note that  by trace inequality \cite[Theorem 1, p. 274]{evans} we have:
\begin{eqnarray}\label{eq:an4}
\int_{\partial \Omega} (\Theta^{-})^2dS\le c_2 \left( \int_{\Omega} |\nabla \Theta^{-}|^2dA+ \int_{\Omega} (\Theta^{-})^2dA \right),
\end{eqnarray}
where $c_2$  depends only on $\Omega$ and $\partial \Omega$.
Combining \eqref{eq:an3} and \eqref{eq:an4} we obtain:
\begin{eqnarray}\label{eq:an5}
q c_1c_2 \left( \int_{\Omega} |\nabla \Theta^{-}|^2dA+ \int_{\Omega} (\Theta^{-})^2dA \right)\ge \int_{\Omega} |\nabla \Theta^{-}|^2dA+h g^{\prime}(0)  \int_{\Omega} (\Theta^{-})^2dA 
\end{eqnarray}
Given normalizations of functions $f$ and $g$ and the domain $\Omega,$ all constants $c_1,c_2$ and $g^{\prime}(0)$ are of order unity.
Thus, we have
\begin{eqnarray}\label{eq:an6}
q   \int_{\Omega} (\Theta^{-})^2dA \ge c_3 \int_{\Omega} |\nabla \Theta^{-}|^2dA+h c_4 \int_{\Omega} (\Theta^{-})^2dA, 
\end{eqnarray}
with $c_3=(1-qc_1c_2)/c_1c_2$ and $c_4=g^{\prime}(0) /c_1c_2$. We note that for $q$ sufficiently small both constants $c_3,c_4$ are positive
and of order one.
As both quantities in the right side of this inequality are positive, this inequality implies that for \eqref{eq:m14} to have a solution we must have:
\begin{eqnarray}\label{eq:an7}
h< c_0 q
\end{eqnarray}
with constant $c_0=1/c_4$ of order one.
Moreover, if the solution of \eqref{eq:m14} exists then
\begin{eqnarray}\label{eq:an7}
\frac{ \int_{\Omega} |\nabla \Theta^{-}|^2dA}{\int_{\Omega} (\Theta^{-})^2dA} < c_0 q. 
\end{eqnarray}
Hence, for any solution of  \eqref{eq:m14} the relative change of $\Theta^-$ over the cross section is negligibly small when $0<q\ll 1$,
that is $\Theta^{-}$ is asymptotically constant.
Therefore,
\begin{eqnarray}\label{eq:an8}
\bar \Theta^- =\frac{1}{|\Omega|} \int_{\Omega} \Theta^- (\xi,\eta) dA \simeq \frac{1}{|\partial \Omega|} \int_{\partial \Omega}  \Theta^-(\xi,\eta)dS,
\end{eqnarray}
where $\simeq$ stands for asymptotically equivalent.


\begin{bibdiv} 
\begin{biblist}

\bib{intro1}{article}{ 
author={MacDonald, D.D},
author={Arthur, H.},
author={Biswas, R.},
 author={ Eklund, K.},
 author={ Hara, N.},
author={ Kakar, G.},
 author={ Kriksunov, L.},
 author={ Liu, C.},
author={ Lvov, S.},
author={Mankowski, J.},
title={Supercritical Water Oxidation Studies: Understanding the Chemistry and Electrochemistry of SCWO Systems},
journal={Final Progress Report of US Army Research Office DAAL 03-92-G-0397 and DAAH 01-93- G-0150},
date={1997},
}

\bib{intro2}{article}{ 
author={Bramlette, T.T.},
author={ Mills, B.E.},
author={ Hencken K.R.},
author={Brynildson, M.E.},
author={ Johnston, S.C.},
author={Hruby, J.M.},
author={ Freemster, H.C},
author={ Odegard, B.C.},
 author={Modell, M.},
 title={ Destruction of DOE/DP Surrogate Wastes with Supercritical Water Oxidation Technology},
 journal={Sandia Report SAND90-8229},
 date={1990},
 }

\bib{intro3}{article}{ 
 author={ Bermejo, M.D.},
author={Cabeza, P. },
 author={ Queiroz, J.P.S. },
  author={ Jim\'enez, C. },
 author={ Cocero, M.J. },
  title={ Analysis of the scale up of a transpiring wall reactor with a hydrothermal flame as a heat source for the supercritical water oxidation},
 journal={J. Supercrit. Fluids},
 volume={56},
  date={2011},
 pages={ 21--32},
 }

\bib{intro4}{article}{ 
author={Cabeza,  P. },
author={  Queiroz, J.P.S.}
author={  Arca, S. },
author={  Jim\'enez, C. },
author={ Guti\'errez, A. },
author={ Bermejo,  M.D.},
author={ Cocero, M.J. },
title={Sludge destruction by means of a hydrothermal flame: optimization of ammonia destruction conditions},
journal= {Chem. Eng. J.},
volume={232},
date= {2013},
pages={1--9},
}

\bib{intro5}{article}{ 
author={Cabeza,  P.},
author={ Bermejo, M.D.}
author={ Jim\'enez, C.}
author={ Cocero, M.J. }
title= {Experimental study of the supercritical water oxidation of recalcitrant compounds under hydrothermal flames using tubular reactors},
journal={Water Res.},
volume={45},
number= {8},
date={2011},
pages= {2485--2495},
}

\bib{intro6}{article}{ 
author={Queiroz, J.P.S. },
author={Bermejo,  M.D. },
author={Mato, F. },
author={ Cocero, M.J.}, 
title={Supercritical water oxidation with hydrothermal flame as internal heat source: efficient and clean energy production from waste},
journal={J. Supercrit. Fluids},
volume={ 96},
 date={2015}, 
 pages={103--113},
 }


\bib{SF}{article}{
author={Schilling, W.},
author={Franck, E.U},
title={Combustion and Diffusion Flames at High Pressures to 2000 bar},
journal={Berichte der Bunsengesellschaft f{\"u}r physikalische Chemie},
date={1988},
Volume={92}, 
number={5},
Pages= {631--636},
}
\bib{MikeRev}{article}{
Author={ Reddy, S.N.},
Author={ Nanda, S.},
Author={ Hegde, U.G.},
Author={ Hicks, M.C.},
Author={ Kozinski, J.A.},
Title={Ignition of hydrothermal flames},
Journal={RSC Advances},
Volume={5},
Year={2015},
Pages={36404--36422},   
}

\bib{Rev1}{article}{
Author={ Augustine, C.}, 
 Author={ Tester, J.W.},
 title={ Hydrothermal flames: from phenomenological experimental demonstrations to quantitative understanding}, 
 journal={J. Supercrit. Fluids},
volume={47},
date={2009},
pages= {415--430},
}

\bib{Grisha_pl}{article}{
author={Sivashinsky, G.I}, 
title={Some developments in premixed combustion modeling}
journal={Proceedings of the Combustion Institute}, 
Volume= {29},
year={2002},
pages={1737--1761},
}




\bib{FK}{book} { AUTHOR = {Frank-Kamenetskii,D. A. }, TITLE
= {Diffusion and heat transfer in chemical kinetics}, PUBLISHER
= { Plenum Press}, ADDRESS = {New York}, YEAR = {1969}, }



\bib{Sem}{article}{ author={Semenov, N.N.}, title={Thermal
theory of combustion and explosion}, Journal={Physics-Uspekhi},
Volume={23}, Issue={3}, year={1940}, pages={251-292}, }


\bib{ZBLM}{book} { AUTHOR = {Zeldovich, Ya. B.}, AUTHOR =
{ Barenblatt, G. I.}, AUTHOR = { Librovich, V. B. }, AUTHOR =
{ Makhviladze, G. M.}, TITLE = {The mathematical theory of combustion
and explosions}, PUBLISHER = {Consultants Bureau {[}Plenum{]}}
ADDRESS = {New York}, YEAR = {1985}, }


\bib{Law}{book}{
author={ Law, C. K. },
title={Combustion Physics},
publisher={Cambridge University Press},
address={Cambridge},
year={2010}
}


\bib{exp}{article}{ author={Hicks, M. C.}, author={Hegde,
U. G.}, author={Kojima, J. J.}, title={Hydrothermal ethanol flames
in co-flow jets}, journal={J. Supercrit. Fluids}, volume={145},
date={2019}, pages={192--200}, }




\bib{Jet1}{article}{ author={Gordon, P. V.}, author={Gotti,
D.J.}, author={Hegde, U.G.}, author={Hicks, M. C.}, author={Kulis,
M.J.}, author={Sivashinsky, G. I.}, title={An elementary model
for autoignition of laminar jets}, journal={Proc. A.}, volume={471},
date={2015}, number={2179}, pages={20150059}, }

\bib{Jet2}{article}{ author={Gordon, P. V.}, author={Hegde,
U. G.}, author={Hicks, M. C.}, author={Kulis, M. J.}, title={On
autoignition of co-flow laminar jets}, journal={SIAM J. Appl. Math.},
volume={76}, date={2016}, number={5}, pages={2081--2098},
}

\bib{kita}{article}{ 
author={ Zhang, Jie },
author={ Zhang,  Lingling},
author={  Men, Chuangshe},
author={  Ren, Mengmeng },
author={ Zhang, Hao} 
author={Lu, Jinling },
title={Ignition of supercritical hydrothermal flames in co-flow jets},
journal={J. Supercrit. Fluids}, volume={188},
date={2022}, pages={105683},
 }


\bib{Jet4}{article}{ author={Gordon, P. V.}, author={Hegde,
U. G.}, author={Hicks, M. C.},
 title={On traveling front of ignition in co-flow laminar reactive jets}, 
journal={SIAM J. Appl.
Math.}, volume={81}, date={2021}, number={1}, pages={47--59},
}

\bib{nist}{webpage}{
    title={https://webbook.nist.gov/chemistry/fluid/},
  }

\bib{hl}{article}{ 
 author={Ju, Y.},
 author={Guo, H.},
 author={Liu, F.},
 author={Maruta, K},
 title={Effects of the Lewis number and radiative heat loss on the bifurcation and extinction of CH4/O2-N2-He flames},
journal={J. Fluid Mech.},
date= {1999}, 
volume= {379},
pages= {165--190},
}


\bib{exp1}{article}{ 
 author={Kojima, Jun J.},
 author={ Hegde, Uday G. },
author={Gotti, Daniel J.},
author={ Hicks, Michael C.},
title={Flame structure of supercritical ethanol/water combustion in a co-flow air stream characterized by Raman chemical analysis},
 journal={J. Supercrit. Fluids}, 
 volume={166},
date={2020}, pages={104995}, 
}

\bib{exp2}{article}{ 
author={ Hegde, Uday },
author={Gotti, Daniel },
author={ Hicks, Michael },
title={The transition to turbulence of buoyant near-critical water jets},
journal={J. Supercrit. Fluids}, volume={95},
date={2014}, pages={195--203},
 }

 
\bib{Jet3}{article}{ author={Gordon, P. V.}, author={Hegde,
U. G.}, author={Hicks, M. C.}, title={An elementary model for
autoignition of free round turbulent jets}, journal={SIAM J. Appl.
Math.}, volume={78}, date={2018}, number={2}, pages={705--718},
}

\bib{GMN2020}{article}{
   author={Gordon, P. V.},
   author={Moroz, V.},
   author={Nazarov, F.},
   title={Gelfand-type problem for turbulent jets},
   journal={J. Differential Equations},
   volume={269},
   date={2020},
   number={7},
   pages={5959--5996},
   }


\bib{Vega}{article}{ author={Vega, Jos\'{e} M.},
title={Travelling wavefronts of reaction-diffusion equations in
   cylindrical domains},
  journal={Comm. Partial Differential Equations},
  volume={18},
  date={1993},
  number={3-4},
  pages={505--531},
  }


\bib{Heinze}{article}{
   author={Heinze, Steffen},
   title={A variational approach to travelling waves},
   journal={Max-Planck-Institute f\"ur Mathematik in den Naturwissenschaften Leipzig Preprints},
      volume={85},
   date={2001},}
   
   
   

\bib{volpert2}{article}{
   author={Apreutesei, Narcisa},
   author={Volpert, Vitaly},
   title={Reaction-diffusion waves with nonlinear boundary conditions},
   journal={Netw. Heterog. Media},
   volume={8},
   date={2013},
   number={1},
   pages={23--35},
%   issn={1556-1801},
%   review={\MR{3043927}},
%   doi={10.3934/nhm.2013.8.23},
}


\bib{kyed}{article}{
   author={Kyed, Mads},
   title={Existence of travelling wave solutions for the heat equation in
   infinite cylinders with a nonlinear boundary condition},
   journal={Math. Nachr.},
   volume={281},
   date={2008},
   number={2},
   pages={253--271},
%   issn={0025-584X},
%   review={\MR{2387364}},
%   doi={10.1002/mana.200710599},
}


\bib{muratov}{article}{
   author={Muratov, C. B.},
   author={Novaga, M.},
   title={Front propagation in infinite cylinders. I. A variational
   approach},
   journal={Commun. Math. Sci.},
   volume={6},
   date={2008},
   number={4},
   pages={799--826},
%   issn={1539-6746},
%   review={\MR{2511694}},
}

\bib{freidlin}{article}{
   author={Freidlin, Mark},
   author={Spiliopoulos, Konstantinos},
   title={Reaction-diffusion equations with nonlinear boundary conditions in
   narrow domains},
   journal={Asymptot. Anal.},
   volume={59},
   date={2008},
   number={3-4},
   pages={227--249},
%   issn={0921-7134},
%   review={\MR{2450360}},
}



\bib{Fife}{article}{
   author={Fife, Paul C.},
   author={McLeod, J. B.},
   title={The approach of solutions of nonlinear diffusion equations to
   travelling front solutions},
   journal={Arch. Rational Mech. Anal.},
   volume={65},
   date={1977},
   number={4},
   pages={335--361},
%   issn={0003-9527},
%   review={\MR{0442480}},
%   doi={10.1007/BF00250432},
}


\bib{evans}{book}{
   author={Evans, Lawrence C.},
   title={Partial differential equations},
   series={Graduate Studies in Mathematics},
   volume={19},
   edition={2},
   publisher={American Mathematical Society, Providence, RI},
   date={2010},
%   pages={xxii+749},
%   isbn={978-0-8218-4974-3},
%   review={\MR{2597943}},
%   doi={10.1090/gsm/019},
}




\end{biblist}
\end{bibdiv}
\end{document}